\magnification = \magstep0
\hsize=18.61truecm  \hoffset=-1.2truecm  \vsize=23.0truecm  \voffset=0.0truecm

\font\sc=cmr9
\font\scit=cmti9
\font\ssc=phvr   scaled 700 
\font\sscap=phvb scaled 700 
\font\cap=phvb scaled 900
\font\sbb=cmbx10 scaled 940

\font\titl=phvb  scaled 2000
\font\stitl=phvb  scaled 1980
\font\au=phvr    scaled 1200
\font\sic=phvro  scaled 700

\def\sglbaselines{\baselineskip=10pt    \lineskip=0pt   \lineskiplimit=0pt}
\def\smlbaselines{\baselineskip=9pt     \lineskip=0pt   \lineskiplimit=0pt}

\def\vsl{\vskip\baselineskip}   \def\vs{\vskip 5pt} 
\parindent=5pt \nopagenumbers

\def\omit#1{\empty}
\def\ba{\kern -4pt}  \def\baa{\kern -6pt}
\parskip = 0pt 
\def\ts{\thinspace} \def\cl{\centerline}
\def\ni{\noindent}  
\def\nhi{\noindent \hangindent=0.9truecm}

\def\makeheadline{\vbox to 0pt{\vskip-30pt\line{\vbox to8.5pt{}\the
                               \headline}\vss}\nointerlineskip}

\def\footnoterule{\kern-3pt \hrule width \fullhsize \kern 2.6pt \vskip 3pt}
\output={\plainoutput}    \pretolerance=10000   \tolerance=10000

\def\gapprox{$_>\atop{^\sim}$} \def\lapprox{$_<\atop{^\sim}$}
          
\newdimen\sa  \def\sd{\sa=.1em \ifmmode $\rlap{.}$''$\kern -\sa$
                               \else \rlap{.}$''$\kern -\sa\fi}
\newdimen\sb  \def\md{\sb=.02em\ifmmode $\rlap{.}$'$\kern -\sb$
                               \else \rlap{.}$'$\kern -\sb\fi}

\def\ss{\ifmmode ^{\prime\prime}$\kern-\sa$ \else $^{\prime\prime}$\kern-\sa\fi}
\def\mm{\ifmmode ^{\prime}$\kern-\sa$ \else $^{\prime}$\kern-\sa \fi}

\sc

\headline{{\bf Nature Letter, 20 January 2011, 000, 000--000 \hfill\null}}

\cl{\null}\vskip -20pt

\noindent{\titl Supermassive black holes do not correlate with dark matter halos of galaxies}

\vskip 10pt

\ni {\au John Kormendy\footnote{$^1$}{\ssc\baa Department of Astronomy, University of Texas at Austin,
                                               1 University Station, Austin, TX 78712-0259, USA}$^{\kern -1pt,2,3}$ \&
Ralf Bender\footnote{$^2$}{\ssc\baa Max-Planck-Institut f\"ur Extraterrestrische Physik, 
                                               Giessenbachstrasse, D-85748 Garching-bei-M\"unchen, 
                                               Germany}$^,$
\ba\footnote{$^3$}{\ssc\baa Universit\"ats-Sternwarte, Scheinerstrasse 1, D-81679 M\"unchen , Germany}
}
\vsl


\newdimen\fullhsize
\fullhsize=18.6 cm 
\hsize=9.05 cm
\def\fullline{\hbox to \fullhsize}
\let\lr=L \newbox\leftcolumn
\output={\if L\lr
         \global\setbox\leftcolumn=\columnbox \global \let\lr=R
         \else  \doubleformat \global \let\lr=L\fi
         \ifnum\outputpenalty>-2000 \else \dosupereject\fi}
\def\doubleformat{\shipout\vbox{\makeheadline
         \fullline{\box\leftcolumn\hfil\columnbox}
         \makefootline}
\advancepageno}
\def\columnbox{\leftline{\pagebody}}


\vskip -1pt

\sglbaselines

{\sbb\cap 
Supermassive black holes have been detected in all galaxies that contain bulge components when the galaxies 
observed were close enough so that the searches were feasible.  Together with the observation that bigger black 
holes live in bigger bulges\raise2pt\hbox{1--\ts4}, this has led to the belief that black hole growth and bulge formation 
regulate each other\raise2pt\hbox{5}. That is, black holes and bulges ``coevolve''.  Therefore, reports\raise2pt\hbox{6,7} 
of a similar correlation between black holes and the dark matter halos in which visible galaxies are embedded have profound 
implications.  Dark matter is likely to be nonbaryonic, so these reports suggest that unknown, exotic physics controls 
black hole growth.~Here we show -- based in part on recent measurements\raise2pt\hbox{8} of bulgeless galaxies -- that 
there is almost no correlation between dark matter and parameters that measure black holes unless the galaxy
also contains~a~bulge.  We conclude that black holes do not correlate directly with dark matter.  They do not
correlate with galaxy disks, either\raise2pt\hbox{9,10}.  Therefore black holes coevolve only with bulges.  
This simplifies the puzzle of their coevolution by focusing attention on purely baryonic processes
in the galaxy mergers that make bulges\raise2pt\hbox{11}.
}

\vskip 5pt

      The idea of coevolution was motivated by the observation~that bigger black holes (BHs) live in bulges and elliptical
galaxies that have bigger velocity dispersions $\sigma$ at large radii where stars feel mainly each others' 
gravity and not that of the BH$^{3,4}$.  This correlation was compelling, because its scatter was small, consistent 
with measurement errors.  The reduced $\chi^2$ was 0.79 for the highest-accuracy sample$^{3}$, and
``the intrinsic scatter~[in~BH mass M$_\bullet$ at fixed $\sigma$] is probably less than 0.15 dex.''$^{4}$  The scatter
was so small that $\sigma$ could be used as a surrogate for M$_\bullet$ for many arguments.  More important was 
the implication that a fundamental physical connection between BH and bulge growth awaits discovery, given 
the realization$^{12}$ that even a tiny fraction of the energy produced in BH growth could, if absorbed by protogalactic gas, 
regulate bulge formation.  Small scatter will be important here, too.~Tight correlations motivate a search for underlying physics.
Loose correlations are less compelling: bigger galaxies just tend to be made of bigger galaxy parts.

      The discovery$^{6,7}$ of a similarly tight correlation between $\sigma$ and the circular rotation velocities 
V\lower2pt\hbox{circ} of gas in the outer parts of galaxies, where gravity is controlled by dark matter 
(DM), therefore was taken to imply that DM also regulates BH growth.  In fact, it was suggested$^6$ that the more
fundamental correlation is the one with DM; i.{\ts}e., that dark matter engineers coevolution.

      The proposed BH--DM correlation raised two concerns.  First, it was known that BHs do not correlate with galaxy 
disks$^{9}$, whereas galaxy disks correlate closely with DM$^{13,14}$.  It was not clear how BHs and disks could separately 
correlate with DM without also correlating with each other.  Second, the velocity resolution of some $\sigma$ measurements
was too low to resolve narrow spectral lines; this problem is discussed in the caption to Figure 1.

      If dark matter controls BH growth and bulges are essentially irrelevant, then V\lower2pt\hbox{circ} should 
correlate tightly with $\sigma$ even in galaxies that do not have bulges.  Figure 1 performs this test.

\vskip 50pt
\cl{\null}
\vskip 10pt
  
      Figure 1 updates the plot that was used to claim$^6$ a \hbox{BH\ts--{\ts}DM} correlation.  The reliable original
data are shown in black; points measured with low velocity resolution were omitted as documented 
in Supplementary Table 1.   Motivated by the above discussion, we measured$^{8}$ velocity  dispersions in 
six Sc\ts--{\ts}Scd galaxies that have nuclear star clusters (``nuclei'') but essentially no bulges.  
They are shown by red points.  Other color points show additional published data on bulgeless galaxies that 
were measured with enough spectral  dispersion to resolve nuclear $\sigma$.

      Figure 1 shows that bulgeless galaxies (color~points, NGC\ts3198) show only a weak correlation between 
V\lower2pt\hbox{circ} and~$\sigma$.~This is expected, because bigger galaxies tend~to have bigger~nuclei$^{20}$.  
But no tight correlation suggests any more compelling formation physics than the expectation that bigger nuclei 
can be manufactured in bigger galaxies that contain more fuel.  The \phantom{000000000}

\vfill

\includegraphics{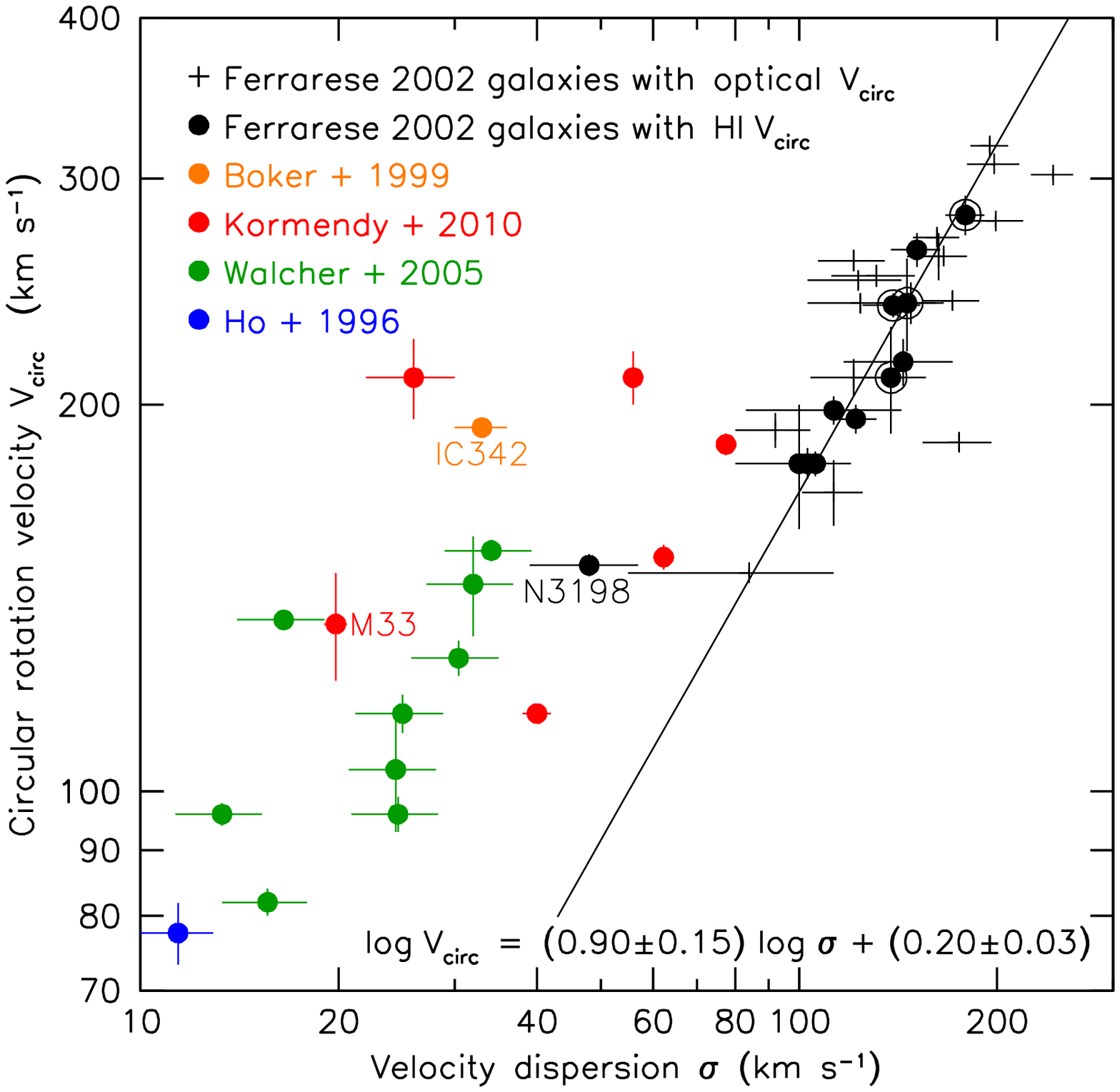}

\ni {\sscap Figure 1 | Outer rotation velocity V\lower2pt\hbox{circ} vs. near-central velocity dispersion~$\sigma$.
The data are listed in the Supplementary Information.  Error bars are~1~sigma.  The original BH--DM correlation\raise2pt\hbox{6} 
is shown in black symbols (circled if the galaxy has a classical bulge) except that points have been omitted if the $\sigma$ measurement
had insufficient velocity resolution. For example, the bulgeless Scd galaxy IC\ts342 (now the orange point, after correction) was 
shown\raise2pt\hbox{6} at $\sigma$ = 77 $\pm$ 12 km s\raise2pt\hbox{-1}, consistent with the black points.  But the 
measurement\raise2pt\hbox{15} had low resolution: the instrumental velocity dispersion
$\sigma$\lower2pt\hbox{instr}\ts=\ts(resolution~FWHM)/2.35 was 61 km s\raise2pt\hbox{-1}, similar to $\sigma$ found in IC 342.  Low 
resolution often results in overestimated $\sigma$.  The same source lists $\sigma$ = 77 km s\raise2pt\hbox{-1} 
for the nucleus of M{\ts}33, which has $\sigma$ = 21 $\pm$ 2 km s\raise2pt\hbox{-1} as measured at high resolution\raise2pt\hbox{16,17}.  
In fact, a high resolution measurement of IC 342 was available\raise2pt\hbox{18}: at $\sigma$\lower2pt\hbox{instr} = 5.5 km s\raise2pt\hbox{-1}, 
$\sigma$ is observed to be 33 $\pm$ 3 km s\raise2pt\hbox{-1} (orange point).  We correct or omit black points if
$\sigma$ \lapprox \ts$\sigma$\lower2pt\hbox{instr}.  We add points (color) for galaxies measured with 
$\sigma$\lower2pt\hbox{instr} $<$ 10 km s\raise2pt\hbox{-1}, i.{\ts}e., high enough resolution to allow measurement of the smallest
dispersions seen in galactic nuclei.  The line (equation at bottom; velocities are in units of 200 km s\raise2pt\hbox{-1}) is a 
symmetric least-squares~fit\raise2pt\hbox{19} 
to the black filled circles minus NGC 3198.  It has $\chi$\raise2pt\hbox{2} = 0.25.  The correlation coefficient is r = 0.95.
This correlation is at least as good as M$_\bullet$ -- $\sigma$.  The correlation for the $+$ points has $\chi$\raise2pt\hbox{2} = 2.6 
and r = 0.77.  In contrast, the correlation for the color points plus NGC 3198 has $\chi$\raise2pt\hbox{2} = 15.7
and r = 0.70.
\lineskip=-8pt \lineskiplimit=-8pt
}

\cl{\null}

\cl{\null}

\vskip 6pt

\vskip 13pt

\cl{\null}

\eject

\noindent scatter is much larger than the measurement errors, $\chi^2$ = 15.7.  Note in particular that
galaxies with $\sigma \simeq 25$ km s$^{-1}$ span almost the complete V\lower2pt\hbox{circ} range for the
color points, 96 -- 210 km s$^{-1}$.

     Our measurements$^{8}$ were made with the 9.2{\ts}m~Hobby-Eberly Telescope and High Resolution Spectrograph; 
$\sigma_{\rm instr}$\ts=\ts8{\ts}km{\ts}s$^{-1}$ reliably resolves the smallest velocity  dispersions seen in 
galactic nuclei.  We easily confirm that $\sigma = 19.8 \pm 0.7$ km s$^{-1}$ in M{\ts}33 and include this 
value in Figure 1.  

     Two properties of our sample deserve emphasis.  First, we observed NGC 5457 = M{\ts}101 and NGC 6946, 
because these are among the biggest bulgeless galaxies (V\lower2pt\hbox{circ} $\simeq$ 210 km~s$^{-1}$).
This is important because Ferrarese concluded$^6$ that ``the V\lower2pt\hbox{circ}--$\sigma$ 
relation \dots~seems to break down below V\lower2pt\hbox{circ} $\sim 150$~km~s$^{-1}$ [in~our notation, so]
 halos of mass smaller than $\sim 5 \times 10^{11}$\ts$M_\odot$ are increasingly less efficient at forming BHs.''  
Our galaxies show that V\lower2pt\hbox{circ} -- $\sigma$ breaks down already at V\lower2pt\hbox{circ} = 210 km s$^{-1}$ 
if the galaxy contains no bulge.
 
     Second, our sample is intentionally biased against galaxies that contain bulges.  We even avoided substantial 
``pseudobulges''; i.{\ts}e., ``fake bulges'' that were made by the internal evolution of galaxy disks$^{21}$ rather 
than by the galaxy mergers that make ``classical bulges''.  The pseudobulge-to-total mass ratios of our galaxies 
are \lapprox {\ts}a few percent; the bulge-to-total ratios are zero.  The relevance of 
pseudobulges is discussed below.  We chose these galaxies because, as noted earlier, {\scit we want to know whether 
DM correlates with BHs in the absence of the component that we know correlates with BHs}.  A study$^{22}$ 
of a large galaxy sample that is not biased against bulges results in similar conclusions: V\lower2pt\hbox{circ} 
correlates weakly with $\sigma$, especially at Hubble types where galaxies contain classical bulges, but the scatter 
is large and ``these results render questionable any attempt to supplant the bulge with the halo as the fundamental 
determinant of the central black hole mass in galaxies.$^{22}$''  Results from this study are included in Figure S3 
in the Supplementary Information.

     Figure 1 shows substantial overlap in V\lower2pt\hbox{circ} between the color points that show little correlation 
and the black filled circles that show a good correlation with $\sigma$.  (The black points shown as error bars are 
for galaxies with only optical rotation curves; they measure V\lower2pt\hbox{circ} less accurately, because they reach 
less far out into the DM halo$^6$.  They show a weak correlation that is not a compelling argument for coevolution.)  
In the overlap range, 180 km s$^{-1}$ \lapprox {\ts}V\lower2pt\hbox{circ} \lapprox \ts220 km s$^{-1}$, galaxies 
participate in the tight V\lower2pt\hbox{circ} -- $\sigma$ shown by the black 
filled circles only if they contain bulges.  Clearly baryons matter to BH growth.  But baryons in a disk are not enough.  
DM~by~itself is not enough.~M{\ts}101 (top-left red point) has a halo that is similar to those of half of the galaxies 
in the tight correlation, but that halo did not manufacture a canonical BH in the absence of a bulge.  
This suggests that bulges, not halos, coevolve with BHs.

      Nevertheless, most black circles in Figure 1 show a correlation whose scatter is consistent with the error bars.  
We need to understand this.  

\hfuzz=20pt

\headline{{\bf Nature, 20 January 2011, 000, 000--000 \hfill\null}\rm 2}

      We suggest that the tight correlation of black points in Figure 1 is a result of the well known conspiracy$^{13,14}$ 
between baryons and DM to make featureless rotation curves with no distinction between the parts that are dominated by
baryonic and nonbaryonic matter.  This possibility was considered and dismissed in reference~(6).  However, it is a 
natural consequence of the observation that baryons make up 17\ts\% of the matter in galaxies$^{23}$ and that, to 
make stars, they need to dissipate inside their halos until they are self-gravitating.  This is enough to engineer 
that V\lower2pt\hbox{circ} is approximately the same for DM halos and for disks embedded in them$^{24,25}$.  
That part of the conspiracy~is not shown by Figure 1 because, absent a bulge, disks reach V\lower2pt\hbox{circ}
at large radii that are not sampled by $\sigma$ measurements of nuclei.  

      Bulges dissipate more than disks.  The consequences are shown in Figures S1 and S2 in the Supplementary 
Information.  Figure S1 shows that V\lower2pt\hbox{circ} for the bulge $\approx$ V\lower2pt\hbox{circ} for~the~halo for 
the two highest-V\lower2pt\hbox{circ} galaxies whose points are circled in Figure\ts1.  Figure S2 shows that the same
equality holds reasonably well, given the uncertainties in rotation curve decomposition, for all 
decompositions that we could find that included a bulge.  It holds in just the V\lower2pt\hbox{circ} range, 
180\ts--\ts260  km s$^{-1}$, where the black circles in Figure\ts1 show a tight correlation.  Because a bulge has 
\hbox{V\lower2pt\hbox{circ} $\sim$ $\sqrt{2}\sigma$,} a correlation like that in Figure 1 is expected from Figure S2. 
{\scit All galaxies that participate in the tight correlation in Figure 1 are included in Figure S2.  And all of them
have bulges or pseudobulges.  We conclude that the correlation is nothing more nor less than a restatement of the 
rotation curve conspiracy for bulges and DM.}  It is a consequence of DM-mediated galaxy formation.  The conceptual leap
to a direct causal correlation between DM and BHs is not required by the data. 

\lineskip=-10pt \lineskiplimit=-10pt

      So far, we have discussed BH correlations indirectly using the assumption that $\sigma$ is a surrogate for BH 
mass.  We now check this assumption and show that it is not valid for most of the black points that define 
the tight correlation in Figure 1.  If $\sigma$ is not a measure of M$_\bullet$, then this further shows that the
correlation is not a consequence of a BH -- DM coevolution.

      In Figure 2, we examine directly the correlations between~$M_\bullet$ and host galaxy properties for galaxies
in which BHs have~been detected dynamically.   All plotted parameters are published elsewhere.  The galaxy sample 
and plotted data are listed in the Supplementary Information of the accompanying~Letter$^{10}$.  The same galaxies
are shown in all panels except: ellipticals do not appear in panel (c) because they have no disk; bulgeless galaxies 
do not appear in panel (a) because they have no bulge; some bulgeless galaxies and pseudobulge galaxies with 
M$_\bullet$ limits do not appear in panel (b) because $\sigma$ is outside the plot range. 

      The top panels correlate M$_\bullet$ with the (a) luminosity and (b) velocity dispersion of the host-galaxy bulge.  
Ellipticals (black) and classical bulges (red) show the (a) good and (b) better correlations that we have come to expect.

      Figure 2(c) shows$^{10}$ that galaxy disks do not correlate with~M$_\bullet$.  
Disks participate in the rotation curve conspiracy: they and their DM halos have similar V\lower2pt\hbox{circ}.
But their masses -- represented here by their K-band luminosities -- cannot be used to predict M$_\bullet$. 

      Figure 2 also distinguishes ``classical bulges'' (red points) and ``pseudobulges'' (blue points).  Classical bulges 
are essentially indistinguishable in structure and parameter correlations from elliptical galaxies (black points).  
We believe that both formed by galaxy mergers (see below).  Pseudobulges are high-density, central components in galaxies 
that superficially \hbox{resemble~--~and} often are mistaken for -- classical bulges but that can be recognized because 
their properties are more disk-like than those of classical bulges.  We now know that this results from fundamentally 
different formation histories.  Complementary to hierarchical clustering$^{26}$, a new aspect of our understanding
of galaxy formation$^{21}$ is that isolated galaxy disks evolve slowly as nonaxisymmetries such as 
bars redistribute angular momentum.  During this process, pseudobulges are grown out of disk material.
Bulge-pseudobulge classifications are listed for all objects in the sample in 
the Supplemental Information of reference (10).  Panels (a) and (b) of Figure 2 illustrate a conclusion from that Letter
which has consequences~here: Pseudobulges show essentially no correlation between M$_\bullet$ and~$\sigma$. 
Baryons do not predict M$_\bullet$ if they are in a pseudobulge. 

     If M$_\bullet$ and $\sigma$ do not correlate for pseudobulges, then $\sigma$ is not a surrogate for M$_\bullet$ in Figure 1,
either.  Bulge classifications for the Figure 1 galaxies are given in the Supplementary Information~of this paper, and two galaxies 
with no published data are classified. 

\vfill\eject

\cl{\null}

\vskip 6.1truein

\includegraphics{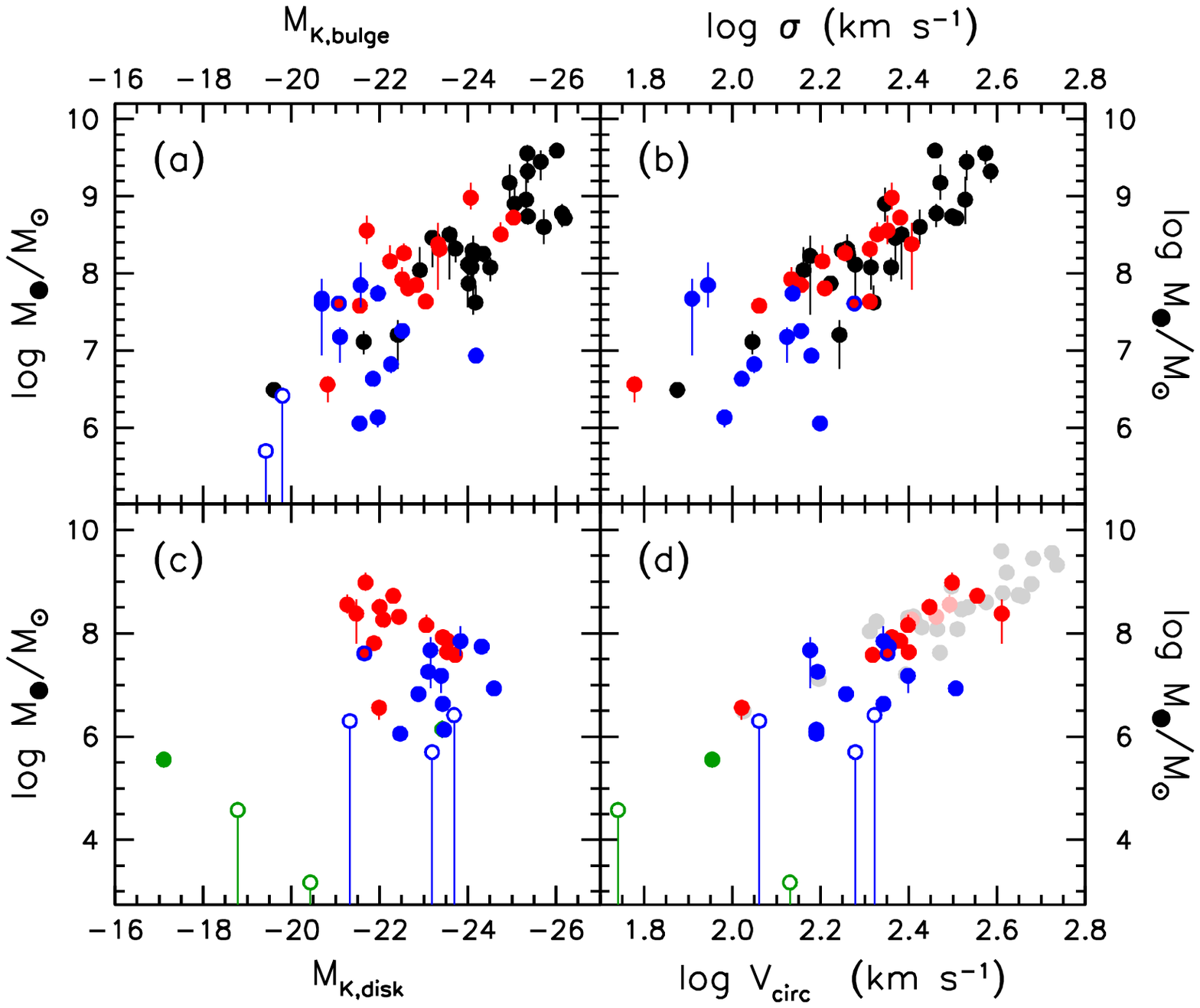}
 
\includegraphics{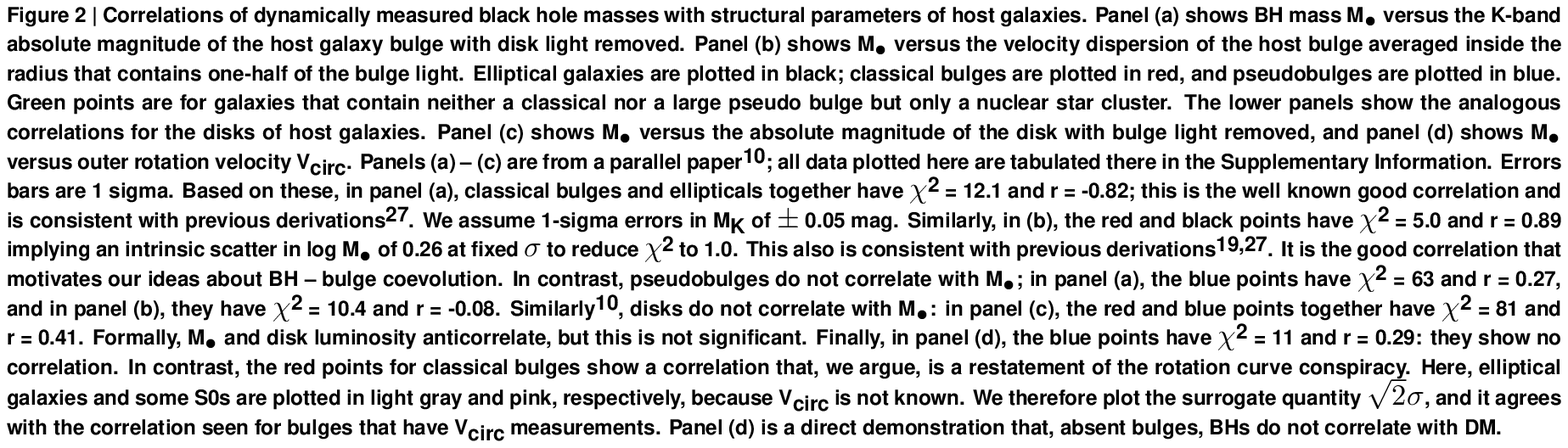}

\vskip 5pt

\headline{{\bf Nature, 20 January 2011, 000, 000--000 \hfill\null}\rm 3}

\noindent We find that only four black circles in Figure 1 correspond to classical bulges, M{\ts}31, 
NGC 2841, NGC 4258, and NGC 7331.  Their points are circled.  The others are for pseudobulges.  For these, the demonstration of a 
tight V\lower2pt\hbox{circ}\ts--\ts$\sigma$ correlation is not a demonstration that DM and BHs correlate.  

      Instead, if V\lower2pt\hbox{circ} correlates with $\sigma$ but $\sigma$ does not measure M$_\bullet$, then this 
supports our conclusion that the correlation results from the rotation curve conspiracy.  Also, the circled points for 
classical bulges agree with the correlation for pseudobulges.  It is implausible to suggest that the correlation for circled points 
is caused by \hbox{BH{\ts}--{\ts}DM} coevolution whereas the identical correlation for the other points has nothing to do with BHs. 

      Finally, Figure 2(d) shows directly that M$_\bullet$ does not correlate with V\lower2pt\hbox{circ} and therefore with
dark matter for pseudobulges. 

      An additional argument is given in \S\ts3 of the Supplementary Information.
If DM V\lower2pt\hbox{circ} predicts M$_\bullet$ independent of baryon content, then the V\lower2pt\hbox{circ} $\sim$ 
1400~km~s$^{-1}$ DM in clusters of galaxies predicts BHs of mass M$_\bullet \sim 7 \times 10^{11}$ M$_\odot$ that
are impossible to hide in well studied clusters such as Coma.

\phantom{0000} \vskip 6.171truein

      Therefore, over the whole range of V\lower2pt\hbox{circ} values associated with dark matter, i.{\ts}e., at least 
50\ts--\ts2000 km s$^{-1}$, Figure 2(d) shows~a correlation with M$_\bullet$ only from 200 -- 400 km s$^{-1}$ plus NGC
7457 at 105 km s$^{-1}$, but only if the galaxy contains a classical bulge.  The bulge correlation can be understood as an 
indirect result of the rotation curve conspiracy.  Even in the above V\lower2pt\hbox{circ} range, there is no correlation 
if the galaxy has only a pseudobulge or disk.  Baryons are not irrelevant.  They are not even sufficient.  To correlate 
with M$_\bullet$, they must be in a classical bulge or elliptical.

      We conclude that BHs do not correlate causally with~DM~halos.   There is no reason to expect that the 
unknown, exotic physics of non-baryonic dark matter directly affects BH growth.   Even DM gravity is not directly
responsible for BH\ts--{\ts}galaxy coevolution.   Rather, that coevolution appears to be as simple as it could be:

       BHs coevolve only with classical bulges and ellipticals. We have a well developed picture of the their formation. 
Hierarchical clustering of density fluctuations in cold dark matter results in frequent galaxy mergers$^{11,26,28}$. 
The products of roughly equal-mass mergers are classical bulges and ellipticals, because progenitor disks get scrambled 
away by dynamical violence$^{11}$.  During this process, gas falls to the center, triggers a burst of star formation$^{29,30}$, 
and builds the high stellar densities that we see in bulges.  This gas may also feed black holes.  In fact, we see a 
correspondence$^{29}$ between mergers-in-progress and quasar-like nuclear activity that builds M$_\bullet$.  Our 
increasingly persuasive picture is that the growth of black holes and the assembly of ellipticals happen together and 
regulate each other$^{12,30}$.  The present results support this picture. 

\vs\vskip 3pt

\smlbaselines

{\frenchspacing

{\ni\sscap Received 15 July; accepted 15 November 2010.} \vs\vskip 3pt

\ssc

\nhi 1.\quad Kormendy, J.~A critical review of stellar-dynamical evidence for black holes in galaxy nuclei.
             in {\sic The Nearest Active Galaxies}. (eds Beckman, J., Colina, L., \& Netzer, H.) 197--218
             (Madrid: Consejo Superior de Investigaciones Cient\'\i ficas, 1993).

\nhi 2.\quad Kormendy, J. \& Richstone, D.~Inward bound\ts--{\ts}The search for supermassive black holes
             in galactic nuclei. {\sic Annu.~Rev.~Astron.~Astrophys.}~{\sscap 33}, 581--624 (1995).

\nhi 3.\quad Ferrarese, L. \& Merritt, D.~A fundamental relation between supermassive black holes and their
              host balaxies.~{\sic Astrophys.~J.}~{\sscap 539}, L9--L12 (2000).

\nhi 4.\quad Gebhardt, K., {\sic et al.}~A relationship between nuclear black hole mass and galaxy velocity 
             dispersion.~{\sic Astrophys.~J.}~{\sscap 539}, L13--L16 (2000).

\nhi 5.\quad Ho, L.~C., Ed.~{\sic Carnegie Observatories Astrophysics Series,
             Volume 1: Coevolution of Black Holes and Galaxies} (Cambridge Univ.~Press, 2004).

\nhi 6.\quad Ferrarese, L.~Beyond the bulge: A fundamental relation between supermassive black holes
             and dark matter halos. {\sic Astrophys.~J.}~{\sscap 578}, 90--97 (2002).

\nhi 7.\quad Baes, M., Buyle, P., Hau, G.~K.~T. \& Dejonghe, H.~Observational evidence for a connection 
              between supermassive black holes and dark matter haloes. 
              {\sic Mon.~Not.~R.~Astron.~Soc.}~{\sscap 341}, L44-L48 (2003).

\nhi 8.\quad Kormendy, J., Drory, N., Bender, R. \& Cornell, M.~E.~Bulgeless giant galaxies
               challenge our picture of galaxy formation by hierarchical clustering.~{\sic
               Astrophys.~J.}~{\sscap 723}, 54--80 (2010).

\nhi 9.\quad Kormendy, J. \& Gebhardt, K.~Supermassive black holes in galactic nuclei.
             in {\sic 20$^{\rm th}$ Texas Symposium on Relativistic Astrophysics}. (eds Wheeler, J. C. \& 
             Martel, H.) 363--381 (AIP, 2001).

\nhi 10.\quad Kormendy, J., Bender, R. \& Cornell, M.~E.~Supermassive black holes
               do not correlate with galaxy disks or pseudobulges. {\sic Nature} \ts{\sscap 000}, 000--000 (2011).

\nhi 11.\quad Toomre, A.~Mergers and some consequences.~in {\sic  Evolution of Galaxies and Stellar 
               Populations} (eds Tinsley, B. M. \& Larson, R. B.) 401--426 (Yale University Observatory, 1977).

\nhi 12.\quad Silk, J. \& Rees, M.~J. Quasars and galaxy formation.
               {\sic Astron. Astrophys.}~{\sscap 331}, L1--L4 (1998).

\nhi 13.\quad van Albada, T.~S. \& Sancisi, R.~Dark matter in spiral galaxies. 
              {\sic Phil.~Trans.~R.~Soc.~London A\/}~{\sscap 320}, 447--464 (1986).

\nhi 14.\quad Sancisi, R. \& van Albada, T.~S.~H I rotation curves of galaxies. in {\sic IAU Symposium 117,
              Dark Matter in the Universe} (eds Kormendy, J. \& Knapp, G. R.) 67--81 (Reidel, 1987).

\nhi 15.\quad Terlevich, E., D\'\i az, A.~I. \& Terlevich, R.~On the behaviour of the IR Ca II triplet
             in normal and active galaxies.~{\sic Mon.~Not.~R.~Astron.~Soc.}~{\sscap 242}, 271--284 (1990).

\nhi 16.\quad Kormendy, J. \& McClure, R.~D.~The nucleus of M{\ts}33.~{\sic Astron.~J.}~{\sscap 105}, 
              1793--1812 (1993).

\nhi 17.\quad Gebhardt, K., {\sic et al.}~M{\ts}33: A galaxy with no supermassive black hole. {\sic
              Astron.~J.}~{\sscap 122}, 2469--2476  (2001).

\nhi 18.\quad B\"oker, T., van der Marel, R.~P. \& Vacca, W.~D.~CO band head spectroscopy of IC 342: 
              Mass and age of the nuclear star cluster.~{\sic Astron.~J.}~{\sscap 118}, 831--842 (1999).

\nhi 19.\quad Tremaine, S., {\sic et al.} The slope of the black hole mass versus velocity dispersion correlation.
             {\sic Astrophys.~J.}~{\sscap 574}, 740--753 (2002).

\nhi 20.\quad B\"oker, T., {\sic et al.}~A Hubble Space Telescope census of nuclear star clusters in 
              late-type spiral galaxies. II. Cluster sizes and structural parameter correlations.
              {\sic Astron.~J.}~{\sscap 127}, 105--118 (2004).

\nhi 21.\quad Kormendy, J. \& Kennicutt, R.~C.~Secular evolution and the formation of pseudobulges
              in disk galaxies.~{\sic Annu.~Rev.~Astron.~Astrophys.}~{\sscap 42}, 603--683 (2004).

\nhi 22.\quad Ho, L. C. Bulge and halo kinematics across the Hubble sequence. 
              {\sic Astrophys.~J.} {\sscap 668}, 94--109 (2007).

\nhi 23.\quad Komatsu, E., {\sic et al.} Five-year Wilkinson Microwave Anisotropy Probe observations: 
               Cosmological interpretation.~{\sic Astrophys.~J.~Suppl.~Ser.}~{\sscap 180}, 330--376 (2009).

\nhi 24.\quad Gunn, J.~E.~Conference summary. in {\sic IAU Symposium 117, Dark Matter in the Universe}
                (eds Kormendy, J. \& Knapp, G. R.) 537--546 (Reidel, 1987).

\nhi 25.\quad Ryden, B.~S. \& Gunn, J.~E.~Galaxy formation by gravitational 
               collapse. {\sic Astrophys.~J.}~{\sscap 318}, 15--31 (1987).

\nhi 26.\quad White, S.~D.~M. \& Rees, M.~J.~Core condensation in heavy halos: A two-stage
               theory for galaxy formation and 
               clustering.~{\sic Mon.~Not.~R.~Astron.~Soc.}~{\sscap 183}, 341--358 (1978).

\nhi 27.\quad G\"ultekin, K., {\sic et al.}~The M--$\sigma$ and M--L relations in galactic bulges,
               and determinations of their intrinsic scatter. {\sic Astrophys.~J.}~{\sscap 698},
               198--221 (2009).

\nhi 28.\quad Springel, V. Simulations of the formation, evolution and clustering of galaxies and quasars.
              {\sic Nature}~\ts{\sscap 435}, 629--636 (2005). 

\nhi 29.\quad Sanders, D.~B., Soifer, B.~T., Elias, J.~H., Madore, B.~F., Matthews, K.,
               Neugebauer, G. \& Scoville, N.~Z.~Ultraluminous infrared galaxies and the
               origin of quasars.~{\sic Astrophys.~J.}~{\sscap 325}, 74--91 (1988). 

\nhi 30.\quad Hopkins, P.~F., Hernquist, L., Cox, T.~J., di Matteo, T., Robertson, B.
               \& Springel, V.~A unified, merger-driven model of the origin of starbursts, 
               quasars, the cosmic X-ray background, supermassive black holes, and galaxy 
               spheroids.~{\sic Astrophys.~J.~Suppl.~Ser.}~{\sscap 163}, 1--49 (2006).

}

\vsl

{\ni\sscap Supplementary Information} {\ssc is linked to the online version of the paper 
                                        at www.nature.com/nature.}

\vsl

\headline{{\bf Nature, 20 January 2011, 000, 000--000 \hfill\null}\rm 4}

{\ni\sscap Acknowledgments} {\ssc We thank Stephane Courteau for making available his surface
photometry of NGC 801 (S.{\ts}I.) and Jenny Greene for helpful comments on the MS.
The Hobby-Eberly Telescope (HET) is a joint project of 
the University of Texas at Austin, Pennsylvania~State~University, Stanford University
Ludwig-Maximilians-Universit\"at Munich, and Georg-August-Universit\"at G\"ottingen. 
The HET is named in honor of its principal benefactors, William P.~Hobby and Robert E.~Eberly.  
This work was supported by the National Science Foundation.}

\vsl

{\ni\sscap Author Contributions} {\ssc\frenchspacing Both authors contributed to the analysis in this paper.
                                                     J.K. wrote most of the text.}

\vsl

{\ni\sscap Author Information} {\ssc Reprints and permissions information is available at
www.nature.com/reprints.  The authors declare no competing financial interests.
Correspondence and requests for materials should be addressed to J.K. (kormendy@astro.as.utexas.edu).

\cl{\null}
\vfill\eject

\noindent{\stitl Supplementary Information}

\headline{{\bf Nature, 20 January 2011, 000, 000--000 \hfill\null}\rm 5}

\vsl

\rm \sglbaselines

      We expand on four issues: \S\ts1 explains the consequences for this paper of the ``conspiracy'' 
{\it in the biggest~disk~galaxies} that bulges, disks, and dark matter halos are arranged in density and radius 
so that their combined, circular-orbit rotation curves are almost flat.  Section 2 
reconstructs Figure 1 for the rotation curve sample used in Section 1 and updates it with newer and more 
accurate measurements.  The result (Fig.~S3) is that the strong V\lower2pt\hbox{circ} -- $\sigma$ correlation 
at high V\lower2pt\hbox{circ} in Figure 1 looks weaker.  Section~3 presents an additional argument 
that dark matter does not predict M$_\bullet$ irrespective of the nature and amount of its baryon content.  
Finally, \S\ts4 documents the data used to construct Figure 1. 

\vs
\cl{\bf 1.~The Conspiracy Between Visible and Dark Matter}
\cl{\bf To Produce Flat, Featureless Rotation Curves}
\vs

\sglbaselines

      Figure S1 illustrates the rotation curve conspiracy$^{13,14,31}$ in the highest- and 
third-highest-V\lower2pt\hbox{circ} galaxies in Figure~1.  In NGC\ts2841, in M{\ts}31, and in high-V\lower2pt\hbox{circ} 
galaxies~in~general, bulges, disks, and DM halos are arranged in radius and density so that their combined rotation 
curves are so flat and featureless that one cannot easily tell which component dominates at which radius.
To understand rotation curves, it is necessary to decompose them into the contributions from each component$^{32}$. 
Rotation curve decompositions like those illustrated in Figure S1 are available for all of the galaxies in Figure 1 
that have H{\ts}I measurements of V\lower2pt\hbox{circ}.  

      A slightly oversimplified paraphrase of the conspiracy is that 
each component has approximately the same maximum V\lower2pt\hbox{circ}.  But they reach these maxima at different 
radii -- the bulge near the center, then the disk (including gas) and then the DM at radii 
that are usually outside the visible part of the galaxy (Figure S1).  The important conclusion~is~this: If
V\lower2pt\hbox{circ,bulge} $\simeq$ V\lower2pt\hbox{circ,halo} (in obvious notation), then there is no 
reason to believe that any V\lower2pt\hbox{circ} -- $\sigma$ correlation that remains at high rotation velocities 
in Fig.~1 is a correlation of $\sigma$ and hence $M_\bullet$ with DM.~It may be no more
than the correlation with bulges that we already know about.

      In particular, three of the four highest-V\lower2pt\hbox{circ} galaxies in Figure 1 contain classical
bulges.  They are, from top to bottom, NGC 2841, M{\ts}31, and NGC 7331.
The fourth is NGC 4565, which has both a big boxy pseudobulge -- i.{\ts}e., an edge-on bar -- and a smaller
``disky'' pseudobulge$^{33}$.  Some of the biggest pseudobulges are consistent with the \hbox{BH-host} galaxy 
correlations in the top panels of Figure~2.

      To rephrase our conclusion: We suggest that any V\lower2pt\hbox{circ}\ts--\ts$\sigma$ correlation that 
remains in Figure 1 is due to the rotation curve conspiracy.  Ferrarese$^6$ considered but dismissed this
possibility. However, we can test our suggestion, because bulges dominate the biggest galaxies and then 
disappear as galaxy luminosity decreases.  If we are correct, then we expect that any correlation in Figure 1 breaks 
down at V\lower2pt\hbox{circ} values where bulges become unimportant.  We show in Figure S2 that this happens
at V\lower2pt\hbox{circ} $\ll$ 200 km s$^{-1}$.

      The key to our test is that the rotation curve conspiracy is not perfect$^{31}$: it works best for 
intermediate-luminosity galaxies, but at large radii, ``rotation
curves rise for faint[er] galaxies, fall for bright[er] ones.''  Rephrased in the language of Figure~S1, 
V\lower2pt\hbox{circ} for the central component(s) is smaller than that of the halo in the smallest galaxies
and larger than that of the halo in the biggest galaxies.  We illustrate this point in Figure S2
and then show why it is relevant to our argument.

\cl{\null}

\centerline{\null}

\vfill

\includegraphics{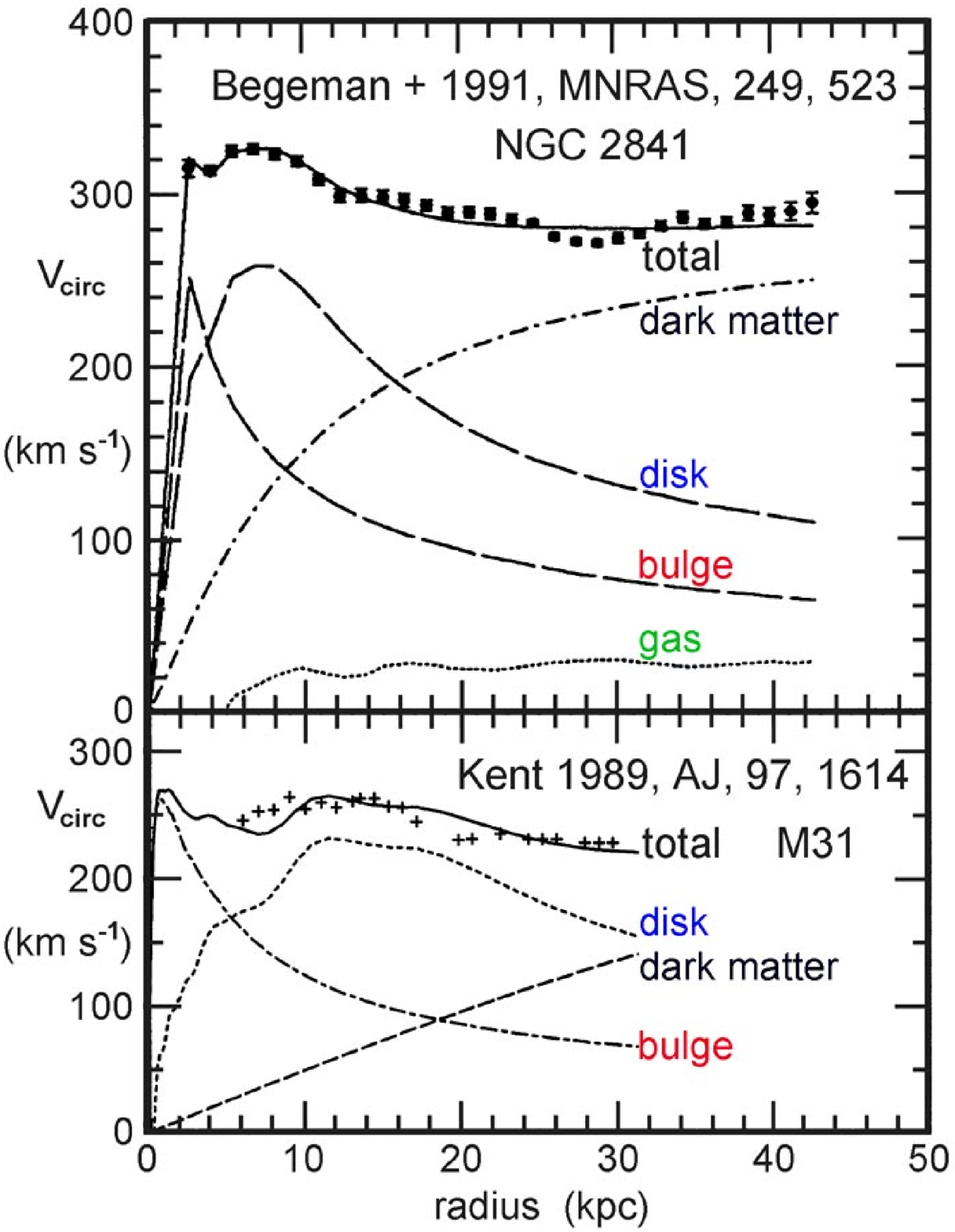}

\ni {\sscap Figure S1 | Rotation curve decompositions for two galaxies in Figure 1 that have classical bulges and
V\lower2pt\hbox{circ} $>$ 200 km s\raise2pt\hbox{-1}.  The standard technique\raise2pt\hbox{32} is to determine
the bulge and disk rotation curves from their radial brightness distributions via the assumption that their 
mass-to-light ratios M/L are constant with radius.  Generally (top) but not always (bottom), H{\ts}I gas is explicitly 
taken into account.  The total rotation curve is the sum in quadrature of its components, i.{\ts}e., 
V\lower2pt\hbox{circ}\raise2pt\hbox{2}(r) = V\lower2pt\hbox{circ,bulge}\raise2pt\hbox{2}
$+$ V\lower2pt\hbox{circ,disk}\raise2pt\hbox{2} $+$ V\lower2pt\hbox{circ,gas}\raise2pt\hbox{2}.  This visible-matter 
rotation curve fits only the central part of the observed rotation curve, some of which is shown by the data points.  
One of the strongest pieces of evidence for dark matter\raise2pt\hbox{34} is that the visible matter cannot explain the flat  
rotation curve at large radii.  Outside the visible matter -- i.{\ts}e., well outside the maxima in V\lower2pt\hbox{circ} 
for visible components -- the total rotation curve should become Keplerian, V\lower2pt\hbox{circ} $\propto$ r\raise2pt\hbox{ -1/2}. 
This is the behavior of the 
bulge rotation curves in Figure S1, because most radii shown are outside most bulge mass.  Because 
the total rotation curve stays flat rather than becoming Keplerian, sufficient dark matter must live at large radii 
so that the new sum in quadrature of V\lower2pt\hbox{circ} for all components fits the data.  There are two
big uncertainties in measuring this dark matter\raise2pt\hbox{32,35}.~Most important are the unknown M/L values.   
A procedure with substantial\raise2pt\hbox{36-41} but not bomb-proof\raise2pt\hbox{32,42-44} 
justification is the ``maximum disk assumption'' in which bulge and disk M/L values are set to the largest values 
that do not over-predict the central rotation curve.  Figure S1 illustrates these maximum disk = minimum halo
models.  The unknown values and radial dependences of M/L exacerbate the second problem, which is that 
the radial density distribution $\rho$(r) of the dark matter is not known.  The usual practice is to assume a functional 
form for $\rho$(r) and then scale the parameters of the halo until it adds up with the visible matter to fit the observed
rotation curve.  In this paper, we use decompositions based on isothermal dark halos or their equivalent\raise2pt\hbox{41}.  
Consistent use of different halo models would lead to somewhat different decompositions quantitatively but the same 
qualitative trends with galaxy luminosity and outer V\lower2pt\hbox{circ}.  In the decompositions illustrated, we 
assume that the rotation curve of the dark matter converges to the flat outer rotation velocity observed.
The conclusion from the decompositions -- and our point in this section -- is that, for each galaxy, the maximum 
V\lower2pt\hbox{circ} is approximately the same for the bulge, the disk, and the dark halo.
\lineskip=-8pt \lineskiplimit=-8pt

\centerline{\null}

}

\eject

\cl{\null} \vskip 3.12truein

\includegraphics{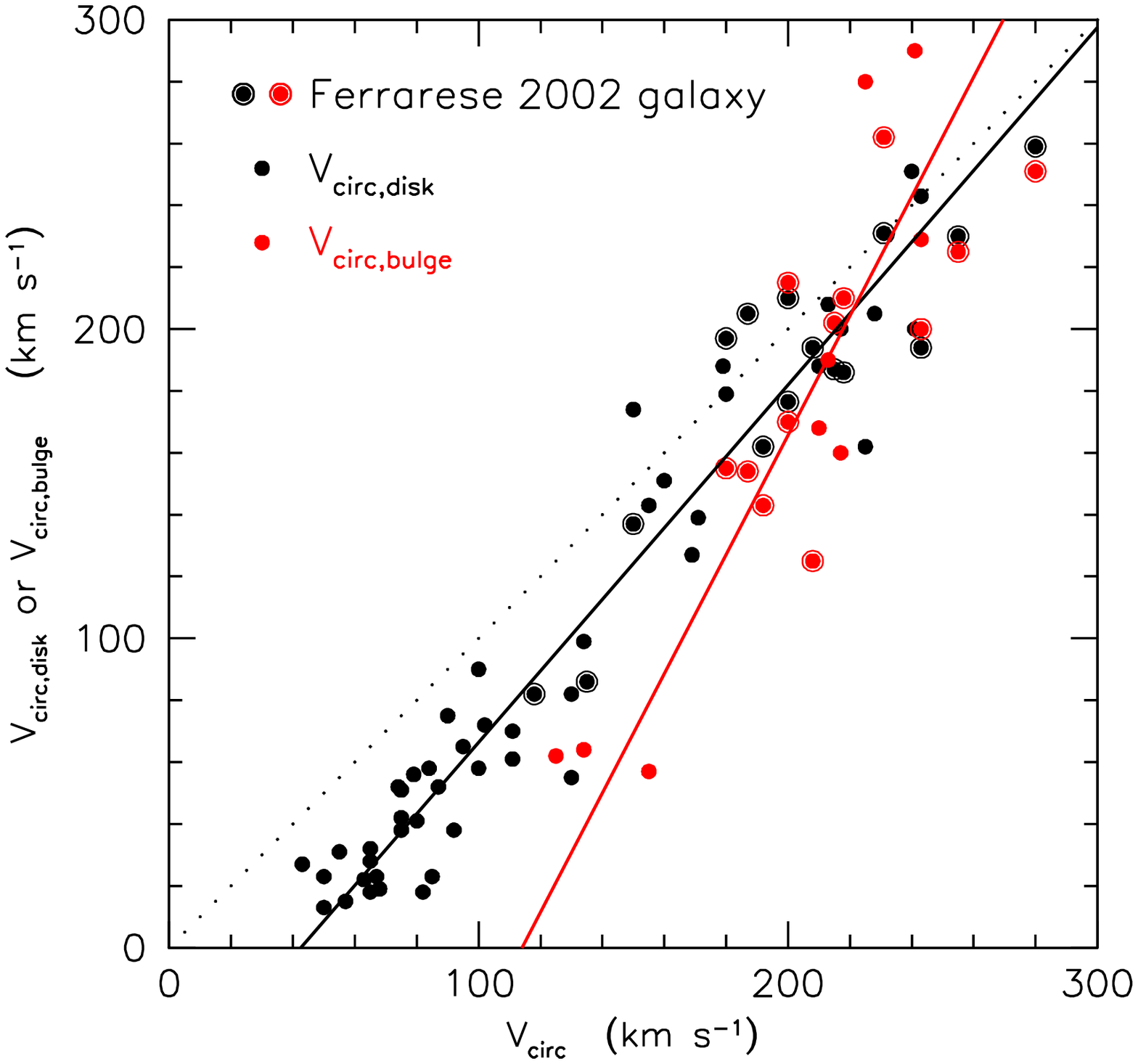}

\ni {\sscap Figure S2 | Maximum rotation velocities of the bulge V\lower2pt\hbox{circ,bulge} (red points)
and disk  V\lower2pt\hbox{circ,disk} (black points) given in bulge-disk-halo decompositions of observed 
rotation curves whose outer rotation velocities are V\lower2pt\hbox{circ}.  Here V\lower2pt\hbox{circ} is
assumed to be the maximum rotation velocity of the dark matter.  The sparse dotted line indicates equality
of the visible matter and dark matter maximum rotation velocities.  Every red point has a corresponding black
point, but many galaxies are sufficiently bulgeless so that only a disk was included in the decomposition
and then the plot shows only a black point.  This is universally true for the smallest galaxies; they 
never contain bulges.  The rotation curve conspiracy happens for galaxies with
V\lower2pt\hbox{circ} $\sim$ 200 km s\raise2pt\hbox{-1}: this is where 
V\lower2pt\hbox{circ,bulge} $\simeq$ V\lower2pt\hbox{circ,disk} $\simeq$ V\lower2pt\hbox{circ} for the
halo.  The correlation for bulges is steeper than that for disks; i.{\ts}e., bulges disappear rapidly
at lower rotation velocities.  The lines are least-squares fits such that the adopted relation y(x) is
the mean of a regression of y on x and one of x on y, where each variable has first been symmetrized 
approximately around its mean\raise2pt\hbox{19}.
The sample is small, but the regressions hint that V\lower2pt\hbox{circ,bulge} tends to be bigger than both
V\lower2pt\hbox{circ,disk} and V\lower2pt\hbox{circ} in the biggest galaxies; this helps to explain why
these galaxies have outward-falling rotation curves\raise2pt\hbox{31}.  Similarly, 
V\lower2pt\hbox{circ,disk} $<$ V\lower2pt\hbox{circ} in the smallest galaxies; this is the other half of
the conspiracy's breakdown\raise2pt\hbox{31}.  In fact, the baryonic
components disappear almost entirely at V\lower2pt\hbox{circ} $\sim$ 45  km s\raise2pt\hbox{-1};
this is an illustration of the well known observation that the smallest galaxies are completely
dominated by dark matter\raise2pt\hbox{41}.  The important point for the present paper is that
V\lower2pt\hbox{circ} \gapprox \ts200  km s\raise2pt\hbox{-1}  is just as much a parameter of the bulge 
as it is a parameter of the dark matter halo.  All galaxies represented by black filled circles in Figure 1 
are also included in Figure S2 (circled points).  Therefore our conclusions apply to these galaxies. 
\lineskip=-8pt \lineskiplimit=-8pt
}

\vskip 5pt

\headline{{\bf Nature, 20 January 2011, 000, 000--000 \hfill\null}\rm 6}

\lineskip=-12pt \lineskiplimit=-12pt

      To check whether the results in Figure S1 are general~and  to see how they
depend on V\lower2pt\hbox{circ}, we plot in Figure S2 the results of rotation curve decompositions
illustrated in the literature$^{32,45-72}$.  The sample is from a study of dark matter scaling laws$^{41}$
augmented by more recent papers.  Selection criteria are as rigorous as practical; only galaxies with H{\ts}I
rotation curves are used, and they need to reach large enough radii to yield reasonably reliable measures
of V\lower2pt\hbox{circ}.  (We can never be completely certain of halo rotation velocities: it is always
possible that they increase again at large radii beyond the reach of present measurements even when  V\lower2pt\hbox{circ}
appears to have flattened out$^{35}$.)  Only maximum-disk or nearly-maximum-disk decompositions and only those that are
based on isothermal halos or equivalent$^{41}$ are used.

\sglbaselines

      Before we interpret Figure S2, there are selection effects in the galaxy sample that we should understand.
The most important one is this:~H{\ts}I gas needs to have been detected to large enough radii to see the rotation
curve flatten.  Many rotation curve decompositions were not used because they do not 
reach large enough radii.  This requirement means that gas must be plentiful.  So late Hubble~types~are~favored.
One result is that the sample of galaxies with bulges -- especially small ones -- is not large.  
However, we were able to find rotation curve decompositions for all of the galaxies shown by filled 
black circles in Figure\ts1; their points are circled in Figure S2.  Also, there are notably many dwarf
galaxies; this is a result of the emphasis put on studying tiny galaxies that are dominated by DM.  It helps
our analysis; it gives us the largest possible dynamic range for a derivation of the underlying
relationship between V\lower2pt\hbox{circ,disk} and V\lower2pt\hbox{circ} (black~straight~line).  
  Finally, we note again that we use maximum disk
decompositions; if disk $M/L$ values were smaller than these maximum values, then V\lower2pt\hbox{circ,disk} would
be correspondingly smaller.  E.{\ts}g., statistical comparisons of vertical velocity dispersions in face-on disks
with vertical scale heights in edge-on disks suggest that $M/L$ may be $\sim$\ts63\ts\% of the maximum-disk value$^{42,43}$.
Then V\lower2pt\hbox{circ,disk} would get $\sim$\ts20\ts\% smaller, V\lower2pt\hbox{circ,bulge} would get very slightly larger,
and V\lower2pt\hbox{circ} for the halo would necessarily remain unchanged.  Our conclusions would be unchanged, too.

      In agreement with previous work$^{31}$, we conclude that the rotation curve conspiracy works best in the biggest
disk galaxies.  Most galaxies with V\lower2pt\hbox{circ} \gapprox \ts200 km s\raise2pt\hbox{-1} 
contain a large bulge, and many of these bulges~are~classical.  Independent of whether they are classical or 
pseudo, V\lower2pt\hbox{circ,bulge} $\simeq$ V\lower2pt\hbox{circ,disk} $\simeq$ V\lower2pt\hbox{circ}.
At V\lower2pt\hbox{circ} $\ll$ 200~km~s\raise2pt\hbox{-1}, bulges become unimportant as V\lower2pt\hbox{circ}
decreases, partly because they are rarer (caution:~the points in Figure~S2 are not necessarily representative)
and partly because V\lower2pt\hbox{circ,bulge} drops below the value for the halo.  But these smaller galaxies
are the ones for which both Figure~1 and previous work$^{22}$ show no tight V\lower2pt\hbox{circ} -- $\sigma$
correlation that argues for BH -- DM coevolution.

\hfuzz=10pt

\pretolerance=15000  \tolerance=15000

      We conclude that there is no compelling evidence for a direct causal correlation between BHs and DM 
beyond what is implied indirectly  by the rotation curve conspiracy.

\vfill\eject

\vs
\cl{\bf 2. Reconstructing the  V\lower2pt\hbox{circ} -- $\sigma$ Correlation}
\vs

\sglbaselines

      Figure S3 reconstructs Figure 1 for all galaxy samples discussed in this paper.~The large 
scatter in the \hbox{V\lower2pt\hbox{circ}\ts--\ts$\sigma$} correlation reported by Ho$^{22}$
is shown by the gray points.  Superposed in black are results for all rotation curve 
decomposition galaxies plotted in Figure S2 for which $\sigma$ measurements are available. 
They include (circled points) all Ferrarese$^6$ galaxies in Figure 1 (black filled 
circles there) that have H{\ts}I rotation curves.  With the best rotation and dispersion 
measurements available now, their scatter is larger than it was in Figure 1 and consistent
with the Ho points.  Classical bulges (red) and pseudobulges (blue) that host detected BHs
are included from Figure\ts2.  They also are consistent with the Ho points. Ho also includes 18 of the
23 Ferrarese galaxies that had optical but no H{\ts}I rotation data (black points 
marked only with error bars in Figure\ts1).   Finally, Figure S3 shows in green the galaxies from
Figure 1 that have virtually no bulge or pseudobulge; the dispersions are those of their stellar
nuclei.  For these, there can be no confusion with any BH -- (pseudo)bulge correlation, so they
remain the best galaxies with which to look for a \hbox{BH\ts--{\ts}DM} correlation.  None is
seen.  Figure S3 therefore supports our conclusion that the rotation curve conspiracy is sufficient
and that a conceptual leap from a BH\ts--{\ts}bulge correlation to a BH\ts--{\ts}DM correlation 
is not compelled by the data.

\vfill

\includegraphics{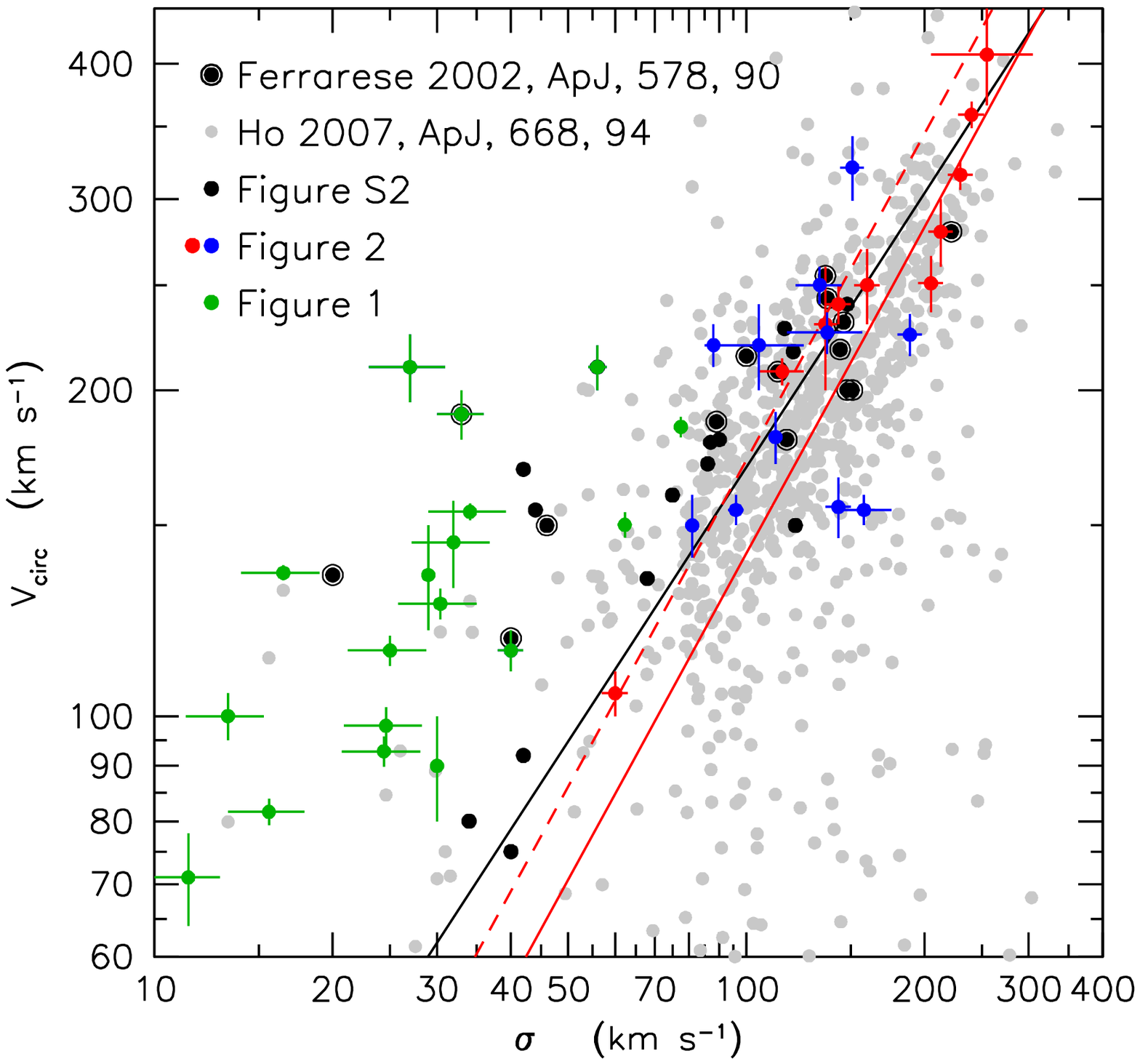}

\ni {\sscap Figure S3 | Reconstruction of Figure 1 for all galaxy samples discussed
in this paper.  Here V\lower2pt\hbox{circ} is the outer rotation velocity of
the galaxy disk, usually measured using observations of H{\ts}I gas, and $\sigma$ is the
central or near-central average velocity dispersion.  Each galaxy is plotted only once;
parameters are taken preferentially from Figure 1, Figure 2, Figure~S2, and Ho\raise2pt\hbox{22}.  All
Ferrarese galaxies with H{\ts}I measurements in Figure 1 (black filled circles there)
are included here but with the best currently available parameters.  Error bars are one sigma.
The black straight line is Ferrarese's fit to her correlation. 
Ho's fit to the light gray points is not shown but is similar.  The red lines are not fits; they
show V\lower2pt\hbox{circ} = $\sqrt{\rm 2}~\sigma$ (lower) and 
V\lower2pt\hbox{circ} = 1.72\ts$\sigma$, respectively.  The latter relation was
derived\raise2pt\hbox{73} by fitting dynamical models with a variety of velocity anisotropies to
x-ray and optical spectroscopy measurements of elliptical galaxies.
An intermediate relation, V\lower2pt\hbox{circ} = 1.52\ts$\sigma$ (not shown) was 
derived\raise2pt\hbox{74} from a dynamical analysis of high-dynamic-range kinematic observations of
ellipticals.  A slope = 1 fit to the black filled circles in Figure 1 (omitting NGC 3198) gives
 V\lower2pt\hbox{circ} = 1.43\ts$\sigma$.

\lineskip=-8pt \lineskiplimit=-8pt
}

\lineskip=-8pt \lineskiplimit=-8pt
}

\eject

\headline{{\bf Nature, 20 January 2011, 000, 000--000 \hfill\null}\rm 7}

\vs
\cl{\bf 3. If Dark Matter Halos Predict BH Masses}
\cl{\bf Independent of their Baryon Content,}
\cl{\bf Then the Halos of Galaxy Clusters}
\cl{\bf Predict Giant Black Holes That Are Not Seen}
\vs

\sglbaselines

      The main text mentions this argument briefly.~We derive~it~here.

      The original $V_{\rm circ}$ -- $\sigma$ correlation$^6$, reproduced to within errors in our Figure 1 (key at the bottom), is
\vskip -4pt
$$
\log{V_{\rm circ}} = (0.84 \pm 0.09)\ts\log{\sigma} + (0.55 \pm 0.19)~, \eqno{(1)}
$$
Substituting for $\sigma$ in the $M_\bullet$ -- $\sigma$ relation$^6$,
$$
M_\bullet = (1.66 \pm 0.32) \times 10^8~M_\odot~\biggl({{\sigma} \over {200~{\rm km~s}^{-1}}}\biggr)^{4.58 \pm 0.52}~, \eqno{(2)}
$$
yields
$$
{{M_\bullet} \over {10^8~M_\odot}} = 0.169~\biggl({V_{\rm circ} \over {200~{\rm km~s}^{-1}}}\biggr)^{5.45} \eqno{(3)}
$$
\vskip 6pt \noindent
in our notation.  As a check on previous arguments, $V_{\rm circ} = 210$ km s$^{-1}$ for M{\ts}101 predicts $M_\bullet \simeq 
2.2 \times 10^7$ $M_\odot$, in conflict with the observed upper limit$^{8}$, $M_\bullet$ \lapprox \ts$(2.6 \pm 0.5) \times
10^6$ M$_\odot$. 
\lineskip=-20pt \lineskiplimit=-20pt

      Rich clusters of galaxies have higher DM $V_{\rm circ}$ values~than~any  galaxy.~They typically have
velocity dispersions $\sigma$\ts$\sim$\ts1000{\ts}km{\ts}s$^{-1}$ and can have velocity dispersions as high as $\sim 2000$ km s$^{-1}$.  
Whether we can use these cluster halos in our argument depends on whether the dark matter is already distributed in the cluster or
whether it is attached only to the galaxies, with the result that the total mass is large but that individual halos~are~not.  
Large-scale simulations of hierarchical clustering show$^{28}$ that, while substructure certainly exists, much of the DM in rich,
relaxed clusters is distributed ``at large'' in the cluster.  In fact, the hierarchical clustering of DM is so nearly scale-free~that$^{75}$
``{\it it is virtually impossible to distinguish [the halo of an individual big galaxy from that of a cluster of galaxies] even
though the cluster halo is nearly a thousand times more massive}'' (emphasis added).  DM halos of mass $10^{15}$ M$_\odot$ are~not rare$^{76}$.  
A cluster of galaxies like Coma$^{77}$ has $\sigma \sim 1000$  km s$^{-1}$ and $V_{\rm circ} \sim \sqrt{2}\ts\sigma \sim 1400$ km s$^{-1}$.
Equation (3) then predicts that 
$$M_\bullet \sim 7 \times 10^{11}~M_\odot~. \eqno(4)$$
Sunk to the center of NGC 4874 or NGC 4889, such a BH would have a sphere-of-influence radius
$r_\bullet \sim G M_\bullet / \sigma^2$\ts$\simeq$\ts34{\ts}kpc\ts=\ts$69^{\prime\prime}$.  This is bigger than the effective radius 
of both galaxies$^{78}$.  But both galaxies have $\sigma \simeq 300$ km{\ts}s$^{-1}$ and normal, linear, and
shallow $\log{\sigma}(\log{r})$ profiles$^{78}$.

      Therefore, if baryons do not matter -- if the hypothesis is that DM makes BHs independent of how baryons are involved 
(as a galaxy, as a group of galaxies, or not at all) -- then Equation (3) predicts unrealistically large M$_\bullet$ for rich
clusters of galaxies.  

      Is this a fair argument?  Does it miss some essential physics that DM requires in order to make BHs?  After all, clusters
are different from individual galaxies, even if both contain DM.  But we emphasize: Any missing physics cannot
involve the formation of bulges and ellipticals, because if those are necessary, then we are back to BH -- bulge coevolution.
The obvious candidate for a missing ingredient is cold gas.  Rich clusters are dominated by X-ray gas.  But the same is true of 
giant ellipticals$^{126}$, and they have manufactured the biggest BHs known.  Both M{\ts}87 and the Coma cluster are dominated 
by stars and hot gas that are only minimally helpful for current BH growth and halos that contain some substructure but that are almost 
self-similar.  Ferrarese (ref.~6) argues that more massive DM halos are more efficient at making BHs.  The most massive halos 
are those of galaxy clusters.  Therefore 
the observation that they do not contain BHs in accord with Equation (3) is a powerful argument that, absent a bulge,
DM does not directly control BH growth.

\vfill\eject

\headline{{\bf Nature, 20 January 2011, 000, 000--000 \hfill\null}\rm 8}

\cl{\null}

\vskip 5truein

\includegraphics{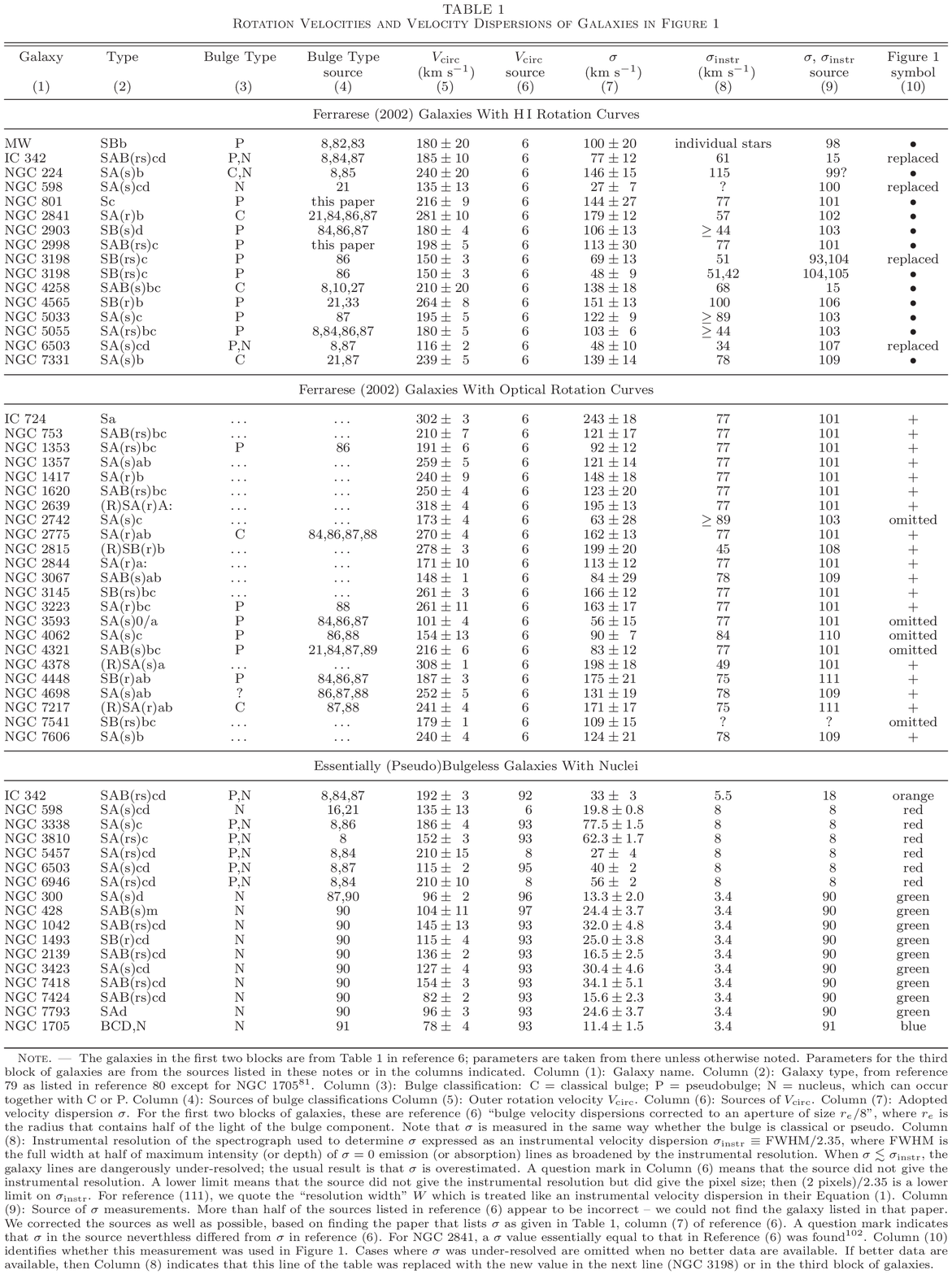}

\cl{\null}

\vfill

\eject

\cl{\null}

\vfill\eject

\vs
\cl{\bf 4. Data Table for Galaxies in Figure 1}
\vs

\headline{{\bf Nature, 20 January 2011, 000, 000--000 \hfill\null}\rm 9}

\lineskip=-20pt \lineskiplimit=-20pt

\sglbaselines

      Table 1 lists the plotted parameters and data sources for all galaxies included
in Figure 1.  The top two blocks of galaxies are taken directly from reference (6) with
as few changes as possible.  If these points are included in Figure 1, they are plotted
in black.  The bottom block of galaxies (color points in Figure 1) includes only objects 
with no classical bulge, with very small pseudobulges (e.{\ts}g., IC 342, NGC 5457, NGC 6946)
or none at all (e.{\ts}g., NGC 598 = M{\ts}33), and with nuclear star clusters whose velocity 
dispersions have been measured with very high velocity resolution (V\lower2pt\hbox{circ}
$<$ 10 km s\raise2pt\hbox{-1}).  In practice, the biggest pseudobulge-to-total luminsity
ratios$^{8}$ included in the bottom group of galaxies in Table 1 are PB/T $\sim$ 0.03.  

      Except for two new bulge-pseudobulge classifications discussed below, all 
data in Table 1 are published.  Column 6 lists the source of V\lower2pt\hbox{circ};
this is taken directly from reference (6) -- where the original source is listed -- for all
galaxies from that paper.  The velocity dispersion $\sigma$ listed by reference (6) is intended
to be an average inside r\lower2pt\hbox{e}/8, where r\lower2pt\hbox{e} is the radius that 
contains half of the light of the bulge.  Some of these values were measured with instrumental
resolution $\sigma$\lower2pt\hbox{instr} $\simeq$ $\sigma$ which is too low (see Figure 1 
caption).  This is documented in Table 1: $\sigma$\lower2pt\hbox{instr} is listed in Column (8)
as reported in the paper that published the measurements (Column 9).  When the resolution was
too poor, the object was ``omitted'' (Column 10) if we had no better $\sigma$ value or ``replaced'' 
with a high-resolution measurement, if we had one.  In the latter case, the galaxy appears again,
either in the next line or in the bottom block of objects, along with the high-resolution $\sigma$ 
measurement and, when necessary, an update on V\lower2pt\hbox{circ}.  Many $\sigma$ sources are 
given incorrectly in reference (6); we list these sources in Column (9), corrected as well as we 
are able (see Table Note for further details).  Column (10) identifies the symbol used for each 
galaxy in Figure 1.

      Readers may worry that $\sigma$ is measured in different ways for the various kinds of
central components included in Figures~1~and~S3.  But in fact, velocity dispersions $\sigma$ 
are defined and measured in exactly the same way for pseudobulges (see below) as for classical bulges. 
In general, the papers that measured r\lower2pt\hbox{e} and $\sigma$ did not distinguish between 
classical and pseudo bulges.  Nuclei are typically only a few arcsec in diameter; their $\sigma$
refers to the whole nucleus.  In one case (NGC 598 = M{\ts}33), measurements with the 
{\scit Hubble Space Telescope} show$^{17}$ that $\sigma$ is independent of radius.  Then it does
not matter whether $\sigma$ is averaged inside r\lower2pt\hbox{e}/8 or not.  For NGC 5457 
and NGC 6946, which contribute more than any other galaxies to our conclusion that DM halos with 
V\lower2pt\hbox{circ} $>$ 200 km s$^{-1}$ do not contain big BHs if they do not also contain classical bulges, 
the nuclear $\sigma$ measurement also includes an equal or larger contribution from the center
of the galaxy's tiny pseudobulge$^{8}$.  In any case, nuclei are so small that, if big DM halos 
manufactured big BHs, then those BHs would easily be revealed by large $\sigma$ values that are
not seen.  And velocity dispersion gradients are shallow enough so that the exact fraction of 
r\lower2pt\hbox{e} inside which $\sigma$ is averaged is not critical. So the comparison of 
colored and black points in Figure 1 is fair.

      Some arguments in the main text depend critically on the distinction$^{10,21}$ between classical
and pseudo bulges in the top block of galaxies\ts--{\ts}the ones that show a tight correlation in 
Fig.~1.  Bulge classifications are less well known for the middle block of galaxies, but these are
also less important, because they do not show a tight V\lower2pt\hbox{circ} -- $\sigma$ correlation.
Table~1 lists the bulge classification in Column (3) and the source of the classification in
Column (4).  For NGC 801 and NGC 2998, published (pseudo)bulge classifications are not available; 
we discuss these galaxies next.

\centerline{\null}\vskip 5pt

      NGC 2998 and NGC 801 are by far the most distant galaxies in the top group in Table 1.
At D = 67 Mpc and 79 Mpc, respectively, they are $\sim$ 4 times farther away than the next most
distant galaxy.  As a result, much less is known about them, and in particular, bulge classifications
have not been published.  However, enough data are available so that we can estimate the bulge type
using the shape of the surface brightness profile.

\cl{\null} \vskip 3.7truein

\includegraphics{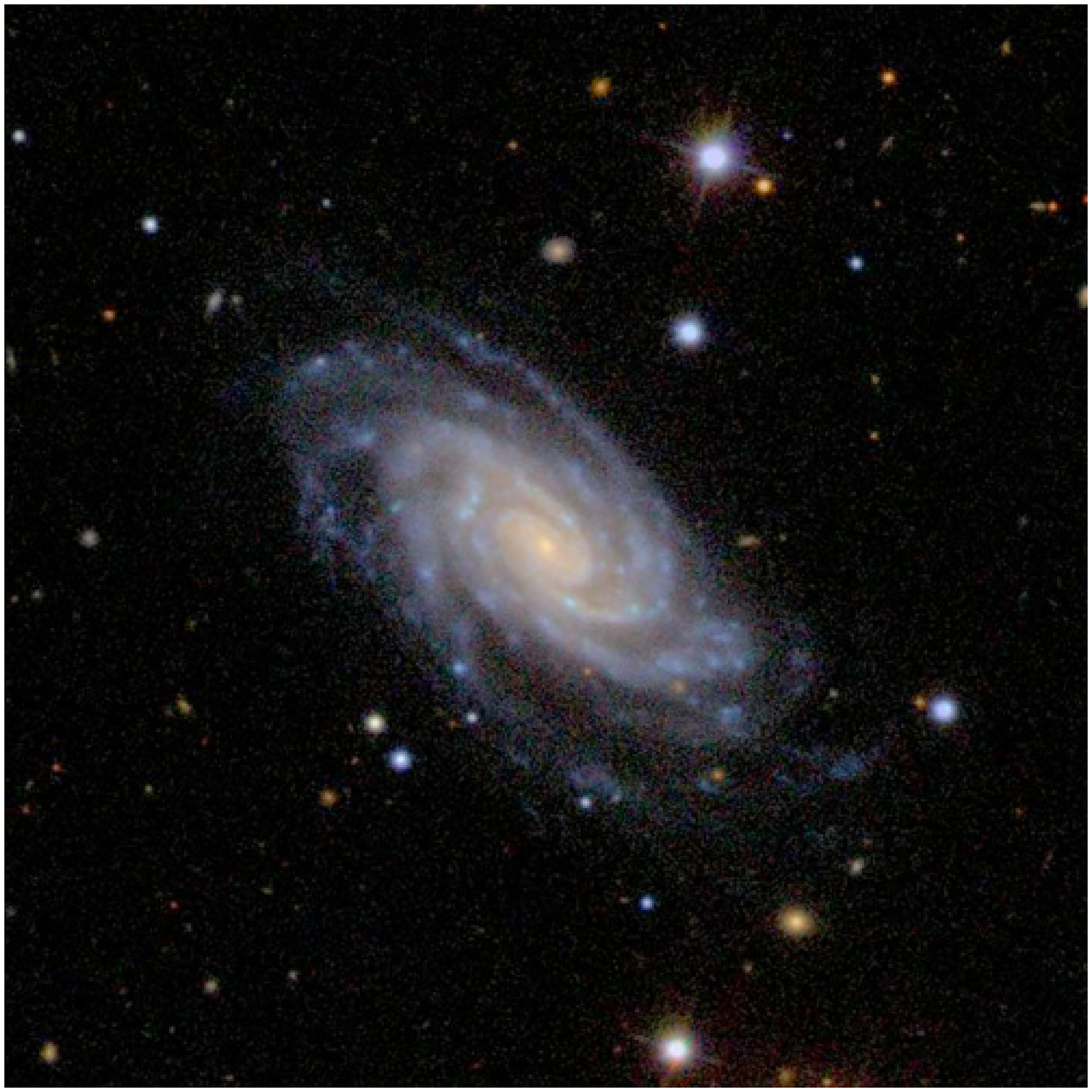}

\ni {\sscap Figure S4 | Color image of the Sc galaxy NGC 2998 constructed using the g-, r-, and i-band images 
from the Sloan Digital Sky Survey (courtesy http://www.wikisky.org).  The galaxy is a close analog of M{\ts}101,
except that it is farther away (D = 67.4 Mpc versus 7.0 Mpc for M{\ts}101) and more inclined.  The total 
K-band absolute magnitude is M\lower2pt\hbox{KT} = -24.2 (compared with -23.7 for~M{\ts}101).  NGC 2998 has 
V\lower2pt\hbox{circ} = 198 $\pm$ 5 km s\raise2pt\hbox{-1} (cf.~210 $\pm$ 15 km s\raise2pt\hbox{-1} 
for M{\ts}101).  We find a pseudobulge-to-total luminosity ratio PB/T = 0.03 $\pm$ 0.01  
(cf.~0.027 $\pm$ 0.008 in M{\ts}101).  The M{\ts}101 parameters are from reference (8).  Error bars are 1 sigma.
\lineskip=-8pt \lineskiplimit=-8pt
}

\vskip 12pt

     NGC 2998 is illustrated in Figure S4.  It is very similar to the Scd galaxy M{\ts}101, which has no classical
bulge at all, but only a nuclear star cluster and a tiny pseudobulge that contributes $0.027 \pm 0.008$ of the
$K$-band light of the galaxy$^8$.  We measured the brightness profile of NGC 2998 to see whether the tiny, bright center 
is a classical or pseudo bulge.  The results are shown in Figure S5.

      As in other normal galaxies, the outer disk has an exponential profile$^{114}$ and the central component 
is well fitted by a S\'ersic function$^{113}$, log I(r) $\propto$ r\raise2pt\hbox{1/n}, where n is the ``S\'ersic index.'' 
The best fit has n = 1.77 $\pm$ 0.15 (one sigma).  This is marginally less than 2.  Much work has 
shown$^{21,84,86-89,115-123}$ that classical bulges have n \gapprox \ts2 whereas most pseudobulges have n $<$ 2.
So our results are most consistent with a pseudobulge, although a classical bulge is not strongly excluded.

      NGC 801 is also classified as an Sc galaxy, but its (pseudo)bulge is brighter than that of NGC 2998.
The galaxy is illustrated in Figure S6.  No HST image is available, but ground-based photometry is collected
into a composite profile in Figure S7.

\vfill

\cl{\null} \vskip 2.7truein

\includegraphics{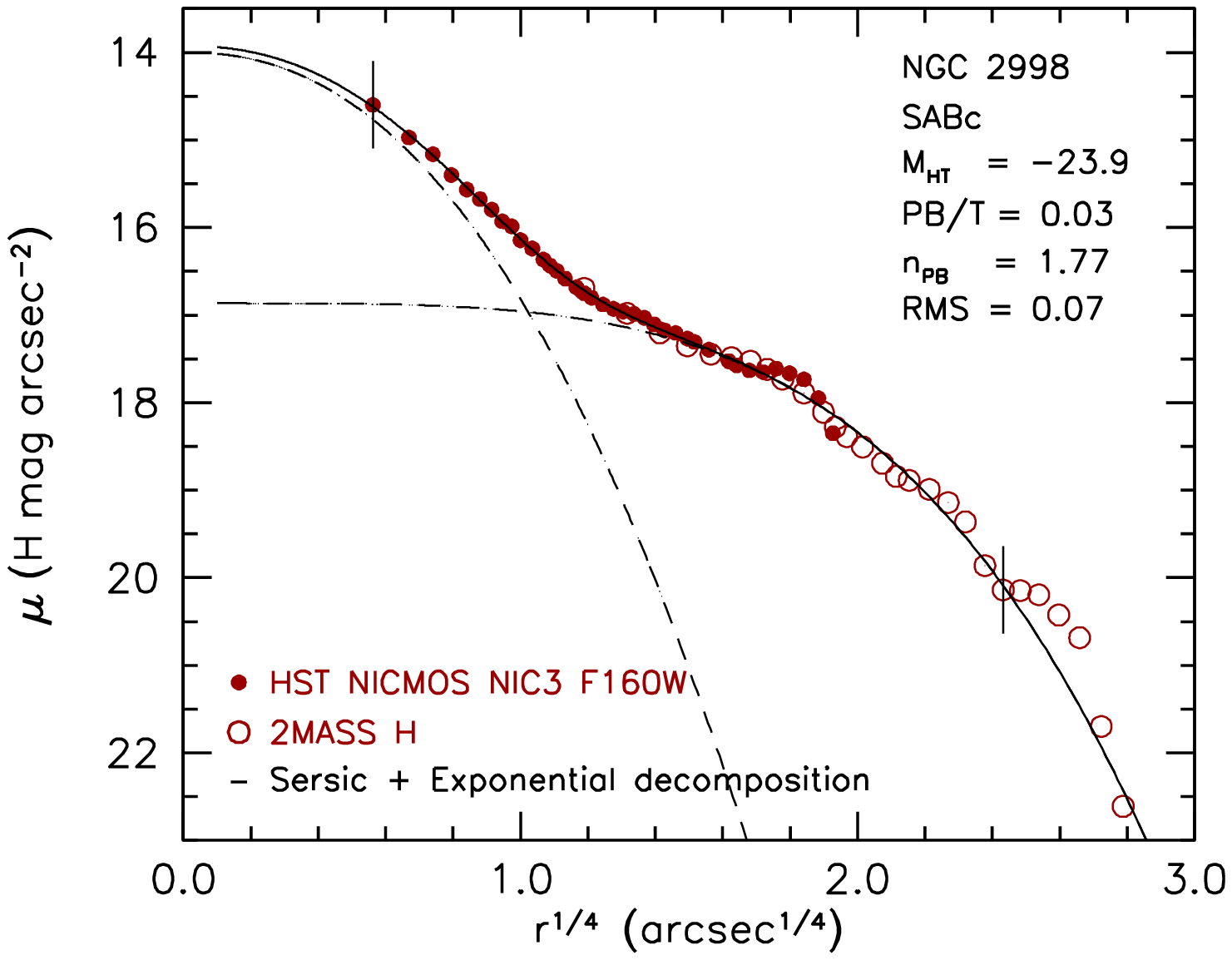}

\ni {\sscap Figure S5 | Major-axis, H-band surface brightness profile~of~NGC~2998.  The filled circles
are our measurements of a Hubble Space Telescope NICMOS image available from the HST Legacy Archive,
http://hla.stsci.edu/hlaview.html.  The open circles are our measurments of the 2MASS H-band image\raise2pt\hbox{112}.
The dashed curves show a decomposition of the profile into a S\'ersic function\raise2pt\hbox{113} (pseudo)bulge
and an exponential disk.  Their intensity sum (solid curve) fits the average profile inside the fit range
(vertical dashes) with an RMS of 0.07 mag arcsec\raise2pt\hbox{-2}.  The S\'ersic index of the
central component is n = 1.77 < 2, so we conclude that this is a pseudobulge (see the text).  The 
pseudobulge-to-total luminosity ratio is tiny, PB/T = 0.03 $\pm$ 0.01 (one sigma).
\lineskip=-8pt \lineskiplimit=-8pt
}

\vskip 8pt

\headline{{\bf Nature, 20 January 2011, 000, 000--000 \hfill\null}\rm 10}

\vskip 3.35truein

\includegraphics{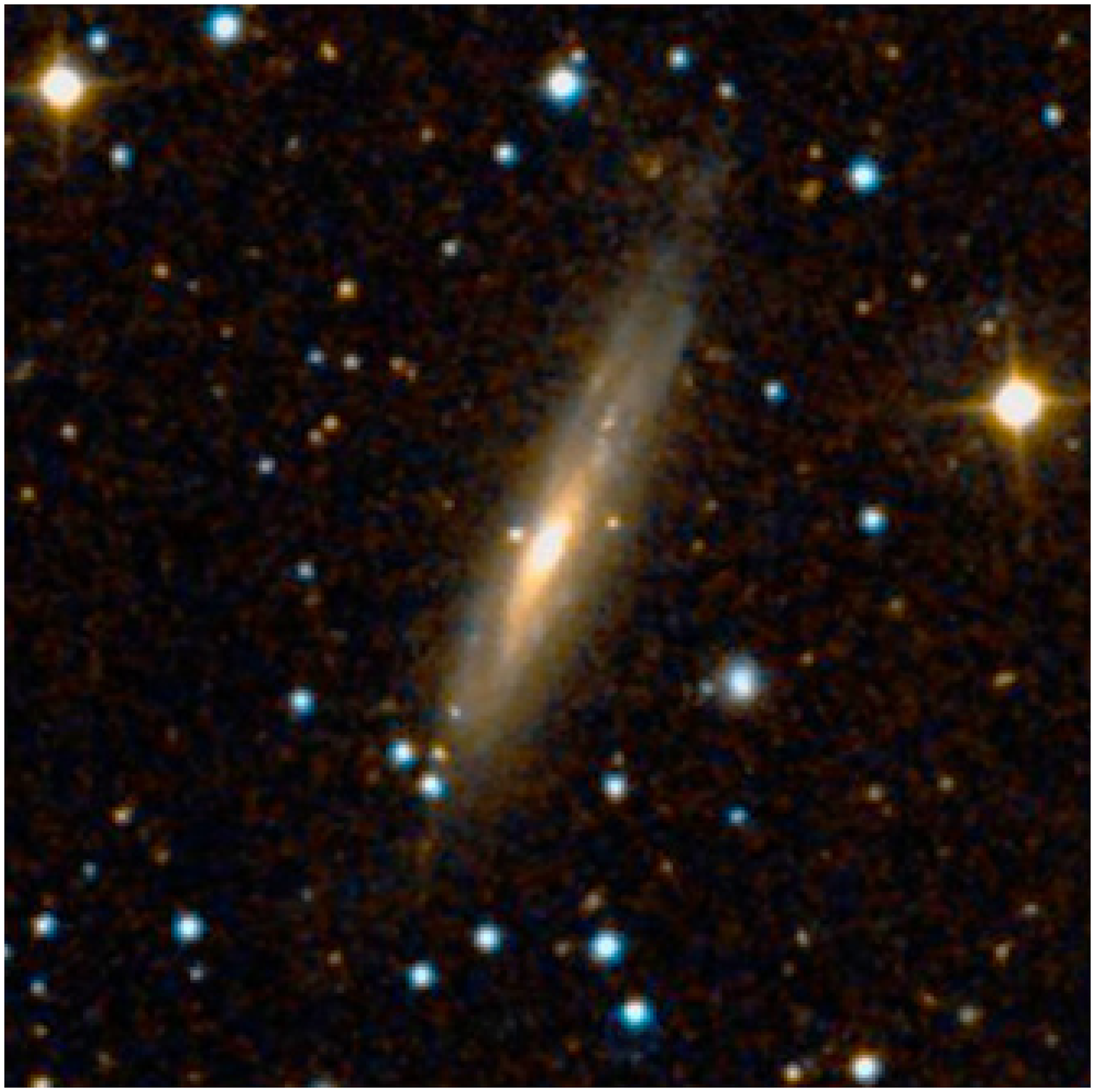}

\ni {\sscap Figure S6 | Color image of the Sc galaxy NGC 801 from the Palomar Observatory and Anglo-Australian
Observatory Digital Sky Survey (courtesy http://www.wikisky.org).  The spatial resolution is poorer than in 
Figure S4; the galaxy is farther away (D = 79 Mpc), and it is almost edge-on.  However, the image correctly suggests 
that the (pseudo)bulge contributes a larger fraction of the galaxy light than it does in NGC 2998.  Otherwise,
NGC 801 is also similar to M{\ts}101: M\lower2pt\hbox{KT} = -25.0 and V\lower2pt\hbox{circ} = 216 $\pm$ 9 km 
s\raise2pt\hbox{-1} (one sigma).
\lineskip=-8pt \lineskiplimit=-8pt
}

\cl{\null}\cl{\null}\cl{\null}

\vskip 2.7truein

\includegraphics{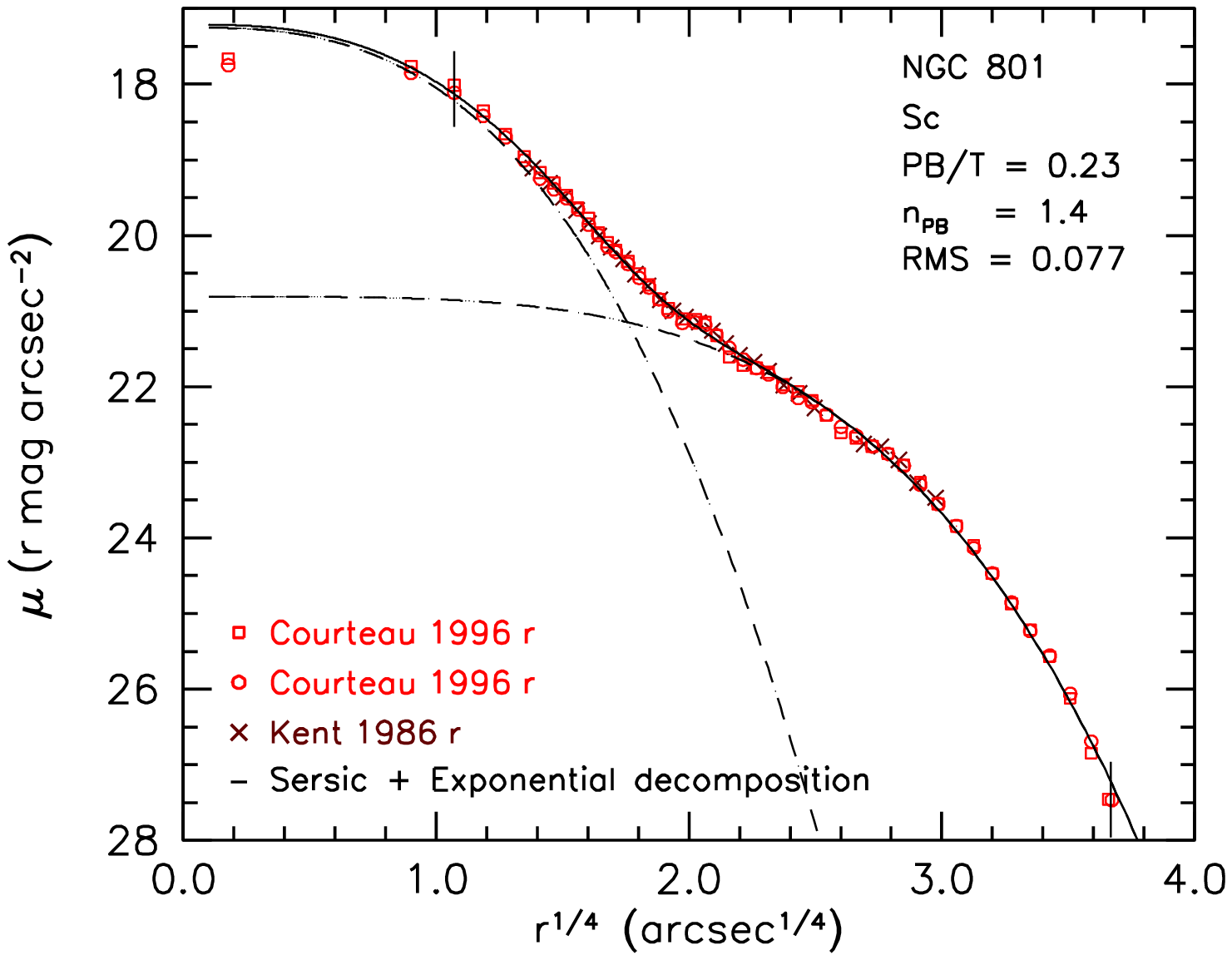}

\ni {\sscap Figure S7 | Major-axis, r-band surface brightness profile\raise2pt\hbox{124,125} of NGC 801.
The dashed curves show a decomposition into a S\'ersic function and an exponential;.  Their sum (solid curve)
fits the profile inside the fit range (vertical dashes) with an RMS of 0.077 mag arcsec\raise2pt\hbox{-2}. 
The S\'ersic index of the central component is n = 1.4 $\pm$ 0.4 (1 sigma).  It contributes $\sim$23\ts\%
of the galaxy light.
\lineskip=-8pt \lineskiplimit=-8pt
}

\vskip 8pt

      Our photometric decomposition (Figure S7) gives a S\'ersic index that is marginally
less than 2.  This is most consistent with a pseudobulge, but a classical bulge is not excluded.
At the large distance of this galaxy, other information such as molecular gas or star formation 
measurements in the central component are not available.  So we cannot apply other classification
criteria.  We provisionally classify the bulge as pseudo but recognize that this is uncertain.

      The pseudobulge classifications of NGC 801 and NGC 2998 are more uncertain than the others
in the top block of  Table~1, because they are based on only one classification criterion.
Also, it is essentially certain that some galaxies have both a classical and a pseudo bulge component.
This is  unlikely in most galaxies in the top block, based on detailed studies.  But it is hard to 
investigate in NGC 801, because that galaxy is far away and because we do not have {\scit Hubble Space Telescope} images.  
(A multicomponent bulge is not likely in NGC 2998: PB/T $\simeq$ 0.03.) 

       However, we emphasize two points.  First, if both of the above classifications were wrong, then 
the tight correlation of black filled circles in Figure 1 would consist of an equal number of classical bulges 
and pseudobulges.  Our arguments in the main text would remain valid.  Second, the criterion that we used to classify NGC 801 
and NGC 2998 was used successfully to classify the pseudobulges that prove to show no $\sigma$ -- M$_\bullet$ correlation$^{10}$.  
If the same classification criterion identifies a similar sample of pseudobulges that show a tight $\sigma$\ts--{\ts}DM 
correlation in Figure\ts1, than that correlation is not due to BH coevolution driven directly by DM.  We discuss 
NGC 801 and NGC 2998 here mainly for completeness.  Our conclusions do not depend on the resulting (pseudo)bulge
classifications.

\vs\vs
\cl{\bf References}
\vs

\smlbaselines

{\frenchspacing

\ssc

\nhi 31.\quad Casertano, S. \& van Gorkom, J.~H.~Declining rotation curves: The end of a conspiracy? 
               {\sic Astron.~J.}~{\sscap 101}, 1231--1241 (1991).

\nhi 32.\quad van Albada, T.~S., Bahcall, J.~N.; Begeman, K. \& Sancisi, R.~Distribution of dark matter 
               in the spiral galaxy NGC 3198. {\sic Astrophys.~J.}~{\sscap 295}, 305--313 (1985).

\nhi 33.\quad Kormendy, J. \& Barentine, J. C. Detection of a pseudobulge hidden inside the ``box-shaped
               bulge'' of NGC 4565. {\sic Astrophys.~J.}~{\sscap 715}, L176--L179 (2010).

\vsl

\nhi 34.\quad Faber, S.~M. \& Gallagher, J. S. Masses and mass-to-light ratios of galaxies.
              {\sic Annu.~Rev.~Astron.~Astrophys.}~{\sscap 17}, 135--187 (1979).

\nhi 35.\quad Lake, G. \& Feinswog, L. The distribution of dark matter in galaxies. I. Models of spiral galaxies.
              {\sic Astron.~J.}~{\sscap 98}, 166--179 (1989).

\nhi 36.\quad Athanassoula, E., Bosma, A. \& Papaioannou, S. Halo parameters of spiral galaxies.
               {\sic Astron.~Astrophys.}~{\sscap 179}, 23--40 (1987).

\nhi 37.\quad Freeman, K. C. Dark matter in galaxies. in {\sic Physics of Nearby Galaxies: Nature or Nurture?}
               (eds Thuan, T. X., Balkowski, C. \& Van, J. T. T.) 201--214 (\'Editions Fronti\`eres, 1992).

\nhi 38.\quad Debattista, V. P. \& Sellwood, J. A. Dynamical friction and the distribution of dark matter in
              barred galaxies. {\sic Astrophys.~J.}~{\sscap 493}, L5--L8 (1998).

\nhi 39.\quad Debattista, V. P. \& Sellwood, J. A. Constraints from dynamical friction on the dark matter 
               content of barred galaxies. {\sic Astrophys.~J.}~{\sscap 543}, 704--721 (2000).

\nhi 40.\quad Weiner, B. J., Sellwood, J. A. \& Williams, T. B. The disk and dark halo mass of the barred balaxy 
               NGC 4123. II. Fluid-dynamical models. {\sic Astrophys.~J.}~{\sscap 546}, 931--951 (2000).

\nhi 41.\quad Kormendy, J. \& Freeman, K.~C.~Scaling laws for dark matter halos in late-type and dwarf 
               spheroidal galaxies. in {\sic IAU Symposium 220, Dark Matter in Galaxies} (eds Ryder, S. D.,
               Pisano, D. J., Walker, M. A. \& Freeman, K. C.) 377--397 (Astronomical Society of the Pacific, 2004).

\nhi 42.\quad Bottema, R. The stellar kinematics of galactic disks. {\sic Astron.~Astrophys.}~{\sscap 275}, 16--36 (1993).

\nhi 43.\quad Bottema, R. The maximum rotation of a galactic disc. {\sic Astron.~Astrophys.}~{\sscap 328}, 517--525 (1997).

\nhi 44.\quad Courteau, S. \& Rix, H.-W. Maximal disks and the Tully-Fisher relation.
              {\sic Astrophys.~J.}~{\sscap 513}, 561--571 (1999).



\nhi 45.\quad Kent, S. M. Dark matter in spiral galaxies. II. Galaxies with H{\ts}I rotation curves.
             {\sic Astron.~J.}~{\sscap 93}, 816--832 (1987).

\nhi 46.\quad Carignan, C. \& Freeman, K. C. DDO 154: A ``dark'' galaxy. {\sic Astrophys.~J.}~{\sscap 332}, L33--L36 (1988).

\nhi 47.\quad Kent, S. M. An improved bulge model for M{\ts}31. {\sic Astron.~J.}~{\sscap 97}, 1614--1621 (1989).

\nhi 48.\quad Jobin, M. \& Carignan, C. The dark side of NGC 3109. {\sic Astron.~J.}~{\sscap 100}, 648--662 (1990).

\nhi 49.\quad Begeman, K. G., Broeils, A. H. \& Sanders, R. H. Extended rotation curves of spiral galaxies: Dark halos
             and modified dynamics. {\sic Mon.~Not.~R.~Astron.~Soc.}~{\sscap 249}, 523--537 (1991).

\nhi 50.\quad Puche, D. \& Carignan, C. H{\ts}I studies of the Sculptor group galaxies. VII. Implications
             on the distribution and nature of dark matter in groups. {\sic Astrophys.~J.}~{\sscap 378}, 487--495 (1991).

\nhi 51.\quad Broeils, A. H. {\sic PhD thesis: Dark and visible matter in spiral galaxies} 1--255 (Rijksuniversiteit te Groningen, 1992)

\nhi 52.\quad Miller, B. W. \& Rubin, V. C. Near-nuclear velocities in NGC 5907: Observations and mass models.
             {\sic Astron.~J.}~{\sscap 110}, 2692--2699 (1995).

\nhi 53\quad Rhee, M.-H. {\sic PhD thesis: A physical basis of the Tully-Fisher relation} 1--137 (Rijksuniversiteit te Groningen, 1996)

\nhi 54.\quad Sofue, Y. The most completely sampled rotation curves for galaxies. 
             {\sic Astrophys.~J.}~{\sscap 458}, 120--131 (1996).

\nhi 55.\quad Verheijen, M. A. W. {\sic PhD thesis: The Ursa Major cluster of galaxies} 1--254 (Rijksuniversiteit te Groningen, 1997)

\nhi 56.\quad Sicotte, V. \& Carignan, C. NGC 5204: A strongly warped Magellanic spiral. II. H{\ts}I kinematics
             and mass distribution. {\sic Astron.~J.}~{\sscap 113}, 609--617 (1997).

\nhi 57.\quad de Blok, W. J. G. \& McGaugh, S. S. The dark and visible matter content of low surface brightness disc galaxies.
             {\sic Mon.~Not.~R.~Astron.~Soc.}~{\sscap 290}, 533-552 (1997).

\nhi 58.\quad van Zee, L., Haynes, M. P., Salzer, J. J. \& Broeils, A. H. A comparative study of star formation 
             thresholds in gas-rich low surface brightness dwarf galaxies. {\sic Astron.~J.}~{\sscap 113}, 1618--1637 (1997).

\nhi 59.\quad Verdes-Montenegro, L., Bosma, A. \& Athanassoula, E. The ringed, warped and isolated galaxy NGC 6015.
             {\sic Astron. Astrophys.}~{\sscap 321}, 754--764 (1997).

\nhi 60.\quad Meurer, G. R., Staveley-Smith, L. \& Killeen, N. E. B. H{\ts}I and dark matter in the windy starburst
             dwarf galaxy NGC 1705. {\sic Mon. Not. R. Astron. Soc.} {\sscap 300}, 705--717 (1998).

\nhi 61.\quad Blais-Ouellette, S., Carignan, C., Amram, P. \& C\^ot\'e, S. Accurate parameters of the mass 
             distribution in spiral galaxies. I. Fabry-Perot observations of NGC 5585. 
             {\sic Astron.~J.}~{\sscap 118}, 2123--2131 (1999).

\nhi 62.\quad C\^ot\'e, S., Carignan, C. \& Freeman, K. C. The various kinematics of dwarf irregular galaxies in nearby
             groups and their dark matter distributions. {\sic Astron.~J.}~{\sscap 120}, 3027--3059 (2000).

\nhi 63.\quad de Blok, W. J. G. \& Bosma, A. High-resolution rotation curves of low surface brightness galaxies.
             {\sic Astron. Astrophys.}~{\sscap 385}, 816--846 (2002).

\nhi 64.\quad Corbelli, E. Dark matter and visible baryons in M{\ts}33. 
             {\sic Mon. Not. R. Astron. Soc.} {\sscap 342}, 199--207 (2003).

\nhi 65.\quad Gentile, G., Salucci, P., Klein, U. \& Granato, G. L. NGC 3741: the dark halo profile from the 
             most extended rotation curve. {\sic Mon. Not. R. Astron. Soc.} {\sscap 375}, 199--212 (2007).

\nhi 66.\quad Noordermeer, E., van der Hulst, J. M., Sancisi, R., Swaters, R. S. \& van Albada, T. S. The mass
             distribution in early-type disc galaxies: Declining rotation curves and correlations with optical
             properties. {\sic Mon. Not. R. Astron. Soc.} {\sscap 376}, 1513--1546 (2007).

\nhi 67.\quad Noordermeer, E. The rotation curves of flattened S\'ersic bulges.
             {\sic Mon. Not. R. Astron. Soc.} {\sscap 385}, 1359--1364 (2008).

\nhi 68.\quad Dicaire, I., {\sic et al.} Deep Fabry-Perot H$\alpha$ observations of NGC 7793: A very extended
             H$\alpha$ disk and a truly declining rotation curve. {\sic Astron.~J.}~{\sscap 135}, 2038--2047 (2008).

\nhi 69.\quad Yoshino, A. \& Ichikawa, T. Colors and mass-to-light ratios of bulges and disks in nearby spiral
             galaxies. {\sic Publ. Astron. Soc. Japan}~{\sscap 60}, 493--520 (2008).

\nhi 70.\quad Sofue, Y., Honma, M. \& Omodaka, T. Unified rotation curve of the Galaxy -- Decomposition into
             de Vaucourleurs bulge, disk, dark halo, and the 9-kpc rotation dip.
             {\sic Publ. Astron. Soc. Japan}~{\sscap 61}, 227--236 (2009).

\nhi 71.\quad Puglielli, D., Widrow, L. M. \& Courteau, S. Dynamical models for NGC 6503 using a Markov chain
             Monte Carlo technique. {\sic Astrophys.~J.}~{\sscap 715}, 1152--1169 (2010).

\nhi 72.\quad Elson, E. C., de Blok, W. J. G. \& Kraan-Korteweg, R. C. The dark matter content of the blue compact
             dwarf NGC 2915. {\sic Mon. Not. R. Astron. Soc.} {\sscap 404}, 2061--2076 (2010).

\nhi 73.\quad Churazov, E., {\sic et al.} Comparison of approximately isothermal gravitational potentials of elliptical
              galaxies based on x-ray and optical data. {\sic Mon. Not. R. Astron. Soc.} {\sscap 404}, 1165--1185 (2010).

\nhi 74.\quad Gerhard, O., Kronawitter, A., Saglia, R. P. \& Bender, R. Dynamical family properties and dark halo 
              scaling relations of giant elliptical galaxies.
              {\sic Astron.~J.}~{\sscap 121}, 1936--1951 (2001).

\headline{{\bf Nature, 20 January 2011, 000, 000--000 \hfill\null}\rm 11}

\nhi 75.\quad Moore, B., Ghigna, S., Governato, F., Lake, G., Quinn, T. \& Stadel, J. Dark matter
              substructure within galactic halos.
              {\sic Astrophys.~J.}~{\sscap 524}, L19--L22 (1999).

\nhi 76.\quad  Faltenbacher, A., Finoguenov, A. \& Drory, N. The halo mass function conditioned on density from the Millennium Simulation: 
               Insights into missing baryons and galaxy mass functions.
               {\sic Astrophys.~J.}~{\sscap 712}, 484--493 (2010).

\nhi 77.\quad Kent, S. M. \& Gunn, J. E. The dynamics of rich clusters of galaxies. I. The Coma Cluster.
              {\sic Astron. J.}~{\sscap 87}, 945--971 (1982).

\nhi 78.\quad Fisher, D., Illingworth, G. \& Franx, M. Kinematics of 13 brightest cluster galaxies.
              {\sic Astrophys.~J.}~{\sscap 438}, 539--562 (1995).


\nhi 79.\quad de Vaucouleurs, G., de Vaucouleurs, A., Corwin, H. G., Buta, R. J., Paturel, G. \& Fouqu\'e, P.
               {\sic Third Reference Catalogue of Bright Galaxies}. (Springer, 1991).

\nhi 80.\quad NASA/IPAC Extragalactic Database (NED), http://nedwww.ipac.caltech.edu/

\nhi 81.\quad Meurer, G. R., Freeman, K. C., Dopita, M. A. \& Cacciari, C. 
              NGC 1705. I.  Stellar populations and mass loss via a galactic wind. 
              {\sic Astron.~J.}~{\sscap 103}, 60--80 (1992).

\nhi 82.\quad Howard, C. D., {\sic et al.} Kinematics at the edge of the Galactic bulge: Evidence for cylindrical rotation.
               {\sic Astrophys.~J.}~{\sscap 702}, L153--L157 (2009).

\nhi 83.\quad Shen, J., Rich, R. M., Kormendy, J., Howard, C. D., De Propris, R. \& Kunder, A.
              Our Milky Way as a pure-disk galaxy -- A challenge for galaxy formation.
               {\sic Astrophys.~J.}~{\sscap 720}, L72--L76 (2010).

\nhi 84.\quad Fisher, D. B., Drory, N. \& Fabricius, M. H. Bulges of nearby galaxies with Spitzer: 
                The growth of pseudobulges in disk galaxies and its connection to outer disks.
                {\sic Astrophys.~J.}~{\sscap 697}, 630--650 (2009).

\nhi 85.\quad Kormendy, J. \& Bender, R. The double nucleus and central black hole of M31.
              {\sic Astrophys.~J.}~{\sscap 522}, 772--792 (1999).

\nhi 86.\quad Fisher, D. B. \& Drory, N. The structure of classical bulges and pseudobulges: The link between pseudobulges 
               and S\'ersic index. {\sic Astron.~J.}~{\sscap 136}, 773--839 (2008).

\nhi 87.\quad Fisher, D. B. \& Drory, N. Bulges of nearby galaxies with Spitzer: 
                Scaling relations in pseudobulges and classical bulges.
                {\sic Astrophys.~J.}~{\sscap 716}, 942--969 (2010).

\nhi 88.\quad  Weinzirl, T., Jogee, S., Khochfar, S., Burkert, A. \& Kormendy, J.
              Bulge n and B/T in high-mass galaxies: Constraints on the origin of bulges in hierarchical models.
              {\sic Astrophys.~J.}~{\sscap 696}, 411--447 (2009).

\nhi 89.\quad Kormendy, J. \& Cornell, M. E. Secular evolution and the growth of pseudobulges in disk galaxies.
              in {\sic Penetrating bars through masks of cosmic dust: The Hubble tuning fork strikes  new note.}
              (eds Block, D. L., Puerari, I., Freeman, K. C., Groess, R. \& Block, E. K.) 261--280 (Kluwer, 2004).

\nhi 90.\quad Walcher, C.~J., van der Marel, R.~P., McLaughlin, D., Rix, H.-W., B\"oker, T.,
             H\"aring, N., Ho, L.~C., Sarzi, M. \& Shields, J.~C.~Masses of star clusters in the 
             nuclei of bulgeless spiral galaxies.~{\sic Astrophys.~J.}~{\sscap 618}, 237--246 (2005).

\nhi 91.\quad Ho, L.~C. \& Filippenko, A.~V.~High-dispersion spectroscopy of a luminous, 
              young star cluster in NGC 1705: Further evidence for present-day formation of 
              globular clusters.~{\sic Astrophys.~J.}~{\sscap 472}, 600--610 (1996).

\nhi 92.\quad Sofue, Y. The most completely sampled rotation curves for galaxies.
             {\sic Astrophys.~J.}~{\sscap 458}, 120--131 (1996).

\nhi 93.\quad HyperLeda database: http://leda.univ-lyon1.fr, see reference 94.

\nhi 94.\quad Paturel, G., Petit, C., Prugniel, Ph., Theureau, G., Rousseau, J., Brouty, M.,
              Dubois, P. \& Cambr\'esy, L.  HYPERLEDA. I. Identification and designation of galaxies.
              {\sic Astron. Astrophys.}~{\sscap 412}, 45--55 (2003).

\nhi 95.\quad van Moorsel, G. A. \& Wells, D. C. Analysis of high-resolution velocity fields -- NGC 6503.
               {\sic Astron.~J.}~{\sscap 90}, 1038--1045 (1985).

\nhi 96.\quad Puche, D., Carignan, C. \& Bosma, A. H{\ts}I studies of the Sculptor Group galaxies. VI. NGC 300.
               {\sic Astron.~J.}~{\sscap 100}, 1468--1476 (1990).

\nhi 97.\quad Epinat, B., Amram, P. \& Marcelin, M.~2008. GHASP: An H$\alpha$ kinematic survey of 203 spiral and 
               irregular galaxies. VII. Revisiting the analysis of H$\alpha$ data cubes for 97 galaxies.
               {\sic Mon. Not. R. Astron. Soc.} {\sscap 390}, 466--504 (2008).

\nhi 98.\quad Merritt, D. \& Ferrarese, L. The M$_\bullet$--$\sigma$ relation for supermassive black holes.
              {\sic Astrophys.~J.}~{\sscap 547}, 140--145 (2001).

\nhi 99.\quad Whitmore, B. C. \& Kirshner, R. P. Velocity dispersions in the bulges of spiral and S0 galaxies. 
              II. Further observations and a simple three-component model for spiral galaxies.
             {\sic Astrophys.~J.}~{\sscap 250}, 43--54 (1981).

\nhi 100.\quad Merritt, D., Ferrarese, L. \& Joseph, C. L. No supermassive black hole in M33?
              {\sic Science}~{\sscap 293}, 1116--1118 (2001).

\nhi 101.\quad Schechter, P. L. New central velocity dispersions for the bulges of 53 spiral and S0 galaxies.
              {\sic Astrophys.~J.~Suppl.~Ser.}~{\sscap 52}, 425--427 (1983).

\nhi 102.\quad Vega Beltr\'an, J. C., Pizzella, A., Corsini, E. M., Funes, J. G., Zeilinger, W. W., Beckman, J. E. \& Bertola, F.
              Kinematic properties of gas and stars in 20 disc galaxies. {\sic Astron. Astrophys.}~{\sscap 374}, 394--411 (2001).

\nhi 103.\quad H\'eraudeau, Ph. \& Simien, F. Stellar kinematical data for the central region of spiral galaxies. I.
              {\sic Astron. Astrophys. Suppl. Ser.}~{\sscap 133}, 317--323 (1998).

\nhi 104.\quad Bottema, R. The stellar kinematics of the spiral galaxies NGC 3198 and NGC 3938.
              {\sic Astron. Astrophys.}~{\sscap 197}, 105--122 (1988).  

\nhi 105.\quad Ho, L. C., Greene, J. E., Filippenko, A. V. \& Sargent, W. L. W. A search for ``dwarf'' Seyfert nuclei. VII.
              A catalog of central stellar velocity dispersions of nearby galaxies.
              {\sic Astrophys.~J.~Suppl.~Ser.}~{\sscap 183}, 1--16 (2009).


\nhi 106.\quad Kormendy, J. \& Illingworth, G. Rotation of the bulge components of disk galaxies.
              {\sic Astrophys.~J.}~{\sscap 256}, 460--480 (1982).

\nhi 107.\quad Bottema, R. The stellar velocity dispersion of the spiral galaxies NGC 6503 and NGC 6340.
              {\sic Astron. Astrophys.}~{\sscap 221}, 236--249 (1989).

\nhi 108.\quad Bottema, R. The stellar velocity dispersion of the spiral galaxies NGC 1566 and NGC 2815.
              {\sic Astron. Astrophys.}~{\sscap 257}, 69--84 (1992).

\nhi 109.\quad H\'eraudeau, Ph., Simien, F., Maubon, G. \& Prugniel, Ph. Stellar kinematic data for the central 
              region of spiral galaxies. II.
              {\sic Astron. Astrophys. Suppl. Ser.}~{\sscap 136}, 509--514 (1999).
 
\nhi 110.\quad Whitmore, B. C., Rubin, V. C. \& Ford, W. K. Stellar and gas kinematics in disk galaxies.
             {\sic Astrophys.~J.}~{\sscap 287}, 66--79 (1984).

\nhi 111.\quad Whitmore, B. C., Kirshner, R. P. \& Schechter, P. L. Velocity dispersions in the bulges of spiral galaxies.
             {\sic Astrophys.~J.}~{\sscap 234}, 68--75 (1979).

\nhi 112.\quad Jarrett, T. H., Chester, T., Cutri, R., Schneider, S. E. \& Huchra, J. P. The 2MASS Large Galaxy Atlas.
              {\sic Astron.~J.}~{\sscap 125}, 525--554 (2003).

\nhi 113.\quad S\'ersic, J.~L. {\sic Atlas de Galaxias Australes}. (Observatorio Astron\'omico, 
               Universidad Nacional de C\'ordoba, 1968)

\nhi 114.\quad Freeman, K. C. 1970, On the disks of spiral and S0 galaxies. 
               {\sic Astrophys.~J.}~{\sscap 160}, 811--830 (1970).

\nhi 115.\quad Courteau, S., de Jong, R.~S. \& Broeils, A.~H.~Evidence for secular evolution in late-type spirals.
               {\sic Astrophys.~J.}~{\sscap 457}, L73--L76 (1996).

\nhi 116.\quad Carollo, C.~M., Stiavelli, M. \& Mack, J. Spiral galaxies with WFPC2. II. The nuclear properties of 40 objects.
               {\sic Astron.~J.}~{\sscap 116}, 68--84 (1998).

\nhi 117.\quad Carollo,~C.~M., Stiavelli, M., de Zeeuw, P.~T., Seigar, M. \& Dejonghe, H. Hubble Space Telescope 
              optical--near-infrared colors of nearby r$^{1/4}$ and exponential bulges.
              {\sic Astrophys.~J.}~{\sscap 546}, 216--222 (2001).

\nhi 118.\quad MacArthur, L.~A., Courteau, S. \& Holtzman, J.~A. Structure of disk-dominated galaxies. I. Bulge/disk parameters, 
               simulations, and secular evolution.
               {\sic Astrophys.~J.}~{\sscap 582}, 689--722 (2003).

\nhi 119.\quad Balcells, M., Graham, A.~W., Dom\'\i nguez-Palmero, L. \& Peletier, R.~F. Galactic bulges from Hubble Space Telescope 
               near-infrared camera multi-object spectrometer observations: The lack of r$^{1/4}$ bulges.
               {\sic Astrophys.~J.}~{\sscap 582}, L79--L82 (2003).

\nhi 120.\quad Fathi, K. \& Peletier, R.~F.~Do bulges of early- and late-type spirals have different morphology?
               {\sic Astron. Astrophys.}~{\sscap 407}, 61--74 (2003).

\nhi 121.\quad Scarlata, C., {\sic et al.}~Nuclear properties of a sample of nearby spiral galaxies from Hubble Space Telescope 
               STIS imaging.
               {\sic Astron.~J.}~{\sscap 128}, 1124--1137 (2004).

\vsl\vsl

\nhi 122.\quad Gadotti, D.~A.~Structural properties of pseudo-bulges, classical bulges and elliptical galaxies: 
               A Sloan Digital Sky Survey perspective
               {\sic Mon. Not. R. Astron. Soc.} {\sscap 393}, 1531--1552 (2009).

\nhi 123.\quad Ganda, K., Peletier, R.~F., Balcells, M. \& Falc\'on-Barroso, J.~The nature of late-type spiral galaxies: 
               Structural parameters, optical and near-infrared colour profiles and dust extinction.
               {\sic Mon. Not. R. Astron. Soc.} {\sscap 395}, 1669--1694 (2009).

\nhi 124.\quad Courteau, S. Deep r-band photometry for northern spiral galaxies.
              {\sic Astrophys.~J.~Suppl.~Ser.}~{\sscap 103}, 363--426 (1996).

\nhi 125.\quad  Kent, S. M. Dark matter in spiral galaxies. I. Galaxies with optical rotation curves.
             {\sic Astron.~J.}~{\sscap 91}, 1301--1327 (1986).

\nhi 126.\quad Kormendy, J., Fisher, D. B., Cornell, M. E. \& Bender, R. Structure and formation of elliptical 
                and spheroidal galaxies. {\sic Astrophys.~J.~Suppl.~Ser.}~{\sscap 182}, 216--309 (2009).

\headline{{\bf Submitted to Nature Letters\hfill\null}\rm 12}

\end